\documentclass{aastex631}
\received{}
\revised{}
\shorttitle{AASTeX v6.31 NICER supersoft}
\shortauthors{Orio et al.}
\graphicspath{{./}{figures/}}

\begin{document}

\title{NICER monitoring of supersoft X-ray sources}

\correspondingauthor{Marina Orio}
\email{orio@astro.wisc.edu, marina.orio@inaf.it}

\author[0000-0003-1563-9803]{M. Orio}
\affiliation{Department of Astronomy, University of Wisconsin 
475 N. Charter Str., Madison, WI, USA}
\affiliation{INAF-Padova, vicolo Osservatorio 5,
35122 Padova, Italy.}


\author{K. Gendreau}
\affil{Center for Exploration and Space Studies (CRESST), NASA/GSFC, Greenbelt, MD 20771, USA}
\affil{NASA Goddard Space Flight Center, Greenbelt, MD 20771, USA}

\author{M. Giese}
\affiliation{Department of Astronomy, University of Wisconsin 
475 N. Charter Str., Madison, WI, USA}

\author[0000-0002-2647-4373]{G. J. M. Luna}
\affil{CONICET-Universidad de Buenos Aires, Instituto de Astronom\'ia y F\'isica del Espacio (IAFE), Av. Inte. G\"uiraldes 2620, 
\\C1428ZAA, Buenos Aires, Argentina}
\affiliation{Universidad de Buenos Aires, Facultad de Ciencias Exactas y Naturales, Buenos Aires, Argentina.}
\affiliation{Universidad Nacional de Hurlingham, Av. Gdor. Vergara 2222, Villa Tesei, Buenos Aires, Argentina}

\author{J. Magdolen}
\affiliation{Advanced Technologies Research Institute, Faculty of Materials Science and Technology in Trnava, Slovak University of Technology in Bratislava, Bottova 25, 917 24 Trnava, Slovakia}
\author[0000-0002-0851-8045]{S. Pei}
\affiliation{Department of Physics and Astronomy, Padova University, vicolo Osservatorio 3, 35122 Padova, Italy}
\affiliation{School of Physics and Electrical Engineering, Liupanshui Normal University, Liupanshui, Guizhou, 553004, China}
\author{B. Sun}
\affiliation{Department of Astronomy, University of Wisconsin 
475 N. Charter Str., Madison, WI, USA}
\affiliation{Department of Physics and Astronomy, University of Missouri, Columbia, MO 65211, USA}
\author{E. Behar}
\affiliation{Department of Physics, Technion, Haifa, Israel}
\author{A. Dobrotka}
\affiliation{Advanced Technologies Research Institute, Faculty of Materials Science and Technology in Trnava, Slovak University of Technology in Bratislava, Bottova 25, 917 24 Trnava, Slovakia}
\author{J. Mikolajewska}
\affiliation{Nicolaus Copernicus Astronomical Center of the Polish Academy
 of Sciences}
\author{Dheeraj R. Pasham}
\affiliation{Kavli Institute for Astrophysics and Space Research, Massachusetts Institute of Technology}
\author[0000-0001-7681-5845]{T. E. Strohmayer}
\affil{Astrophysics Science Division and Joint Space-Science Institute, NASA Goddard Space Flight Center, Greenbelt, MD 20771, USA}


\begin{abstract}
We  monitored  four supersoft sources - two persistent ones, CAL 83 and MR Vel, and
 the recent novae YZ Ret (Nova Ret 2020) and V1674 Her (Nova Her 2021) - with NICER.
 The two persistent SSS were observed with unvaried X-ray
 flux level and spectrum, respectively, 13 and 20 years after the last
 observations. Short period modulations of the supersoft X-ray
 source (SSS) appear where the spectrum of the luminous central source 
 was fully visible
 (in CAL 83 and V1674 Her) and were absent in YZ Ret and MR Vel,
in which  the flux originated in photoionized or shocked
 plasma, while the white dwarf (WD) was not observable. We thus suggest
 that the pulsations occur on, or very close to, the WD surface.  
 The pulsations of CAL 83 were almost unvaried after 15 years,
 including an irregular drift of the $\simeq$67 s 
 period by 2.1 s.  Simulations, including previous
 XMM-Newton data, indicate actual variations in period length within hours,
 rather than an artifact of the variable amplitude of the pulsations.
Large amplitude pulsations with a period of 501.53$\pm$0.30 s were 
always detected in V1674 Her, as long as the SSS was observable.
  This period  seems to be due to rotation of  a highly magnetized WD.
 We cannot confirm the  maximum effective temperature of  ($\simeq$145,000 K) previously inferred for this nova, and discuss the difficulty
 in interpreting its spectrum. The WD appears to
 present two surface zones, one of which does not emit SSS flux.
\end{abstract}


%
\section{Introduction} \label{sec:intro}
Supersoft X-ray sources (SSS) that are also sufficiently luminous at other wavelengths,
 allowing a clear classification, are found to be 
 accreting WDs, undergoing hydrogen burning in
 a shell close to the surface \citep[see, among other articles][and references therein]{Orio2013}. 
 Often, they are transient SSS in post-outburst novae, in which hydrogen burning
 continues near the surface after the bulk of accreted envelope mass has been ejected
 and the WD photosphere has shrunk back to almost pre-outburst radius. 
  The range of effective temperature T$_{\rm eff}$ in the SSS 
 varies from $\simeq$120,000 K to about a million K. Most post-AGB WDs that are still shell-burning hydrogen have larger envelopes on top than post-novae,
 and the average T$_{\rm eff}$ is much lower,
 often peaking only in the ultraviolet \citep[see the objects
 studied by ][]{Corsico2014}. There are a few exceptions,
 mainly PG 1159-type central
 stars of planetary nebulae,  with T$_{\rm eff}>$100,000 K \citep{Lobling2019, Adamczak2012}.
 In the accreting WD of binaries, the burning shell
 is closer to the surface than in the post-AGB stars.
 In novae, convection brings to the surface
 $\beta$+ decaying nuclei that heat the envelope,
 so the average T$_{\rm eff}$ is the highest. 
 Several post-outburst novae have been found to
 be close to the high end of the SSS temperature
 distribution \citep[e.g.][]{Ness2011, Orio2018}.
 A non-negligible group
 of non-novae, semi-steady or recurrent SSS, has been 
 discovered in the Local
 Group \citep[see][]{Orio2010, Orio2013}, in directions of
 low absorbing column of gas. Especially in the Magellanic Clouds,  
 many of these SSS have been identified with WD binaries in which accretion
 occurs onto the WD and hydrogen burning is ignited, but the process occurs at such
 high rate that it does not cause thermonuclear runaways like in novae \citep[see][for
 the theoretical models]{Yaron2005, Starrfield2012, Wolf2013}.
 In this work, we observed again two of the non-nova close binary
 SSS, extending the time line of their observations to over 20 years
 for one of them (MR Vel) and $\simeq$40 years for the other one, CAL 83.
 
  A very interesting characteristic of many SSS is
 the flux modulation with short periods, ranging
 from tens of seconds to $\simeq$ 1 hour. In Table 1, we show a list of measured
 periods of semi-steady SSS and novae in the SSS phase, not including longer
 periods that have also been measured in some cases and have been found to be
 orbital in nature. The periods in the 5-45 minutes range measured in several
 novae, namely 
 V407 Lup, V2491 Cyg and V4743 Sgr have been measured again
 at quiescence when the X-ray flux was $\simeq$ 5 orders of magnitude
 lower and spectrum was less soft, 
 as described in the references in Table 1. For these novae,
 the period has been attributed to the WD spin in a highly magnetized WD, an
 intermediate polar (IP) system. The  period detection at quiescence is 
 understood as due to the accretion curtain/pre-shock material absorbing the X-ray soft radiation of the
 hot polar caps differently as the WD rotates. It is not obvious why the SSS
 emission of the hydrogen burning WD is not homogeneous and why it
 shows a modulation
 with the rotation period. Probably, the WD is hotter at the poles, and/or 
 the burning occurs closer to the surface and lasts longer at the poles, 
 possibly fed by renewed accretion at the end of the outburst \citep[see][for a discussion]{Zemko2016, Aydi2018}. 
 Understanding the mechanism of the modulations of the SSS post-outburst is
 essential to correctly model the outburst and the quiescent life of the systems.

 Also in two non-nova SSS, the periodicity has been linked with the  
 rotation period of the WD. In the WD with the shortest ever detected period of 13.2 s,
 RX J0648.0-4418, a negative period derivative has been measured, and has been interpreted 
 as due to a young, contracting WD in a system containing a subdwarf as close
 binary companion \citep[][and references therein]{Mereg2021}. 
 In CAL83, a period of $\simeq$67 s has also been interpreted as linked with
 rotation. The period appeared to have a drift of about 3 s within short
 timescales, interpreted by \citet{Odendaal2014} as due to an envelope that is not quite synchronized with the rotation of the core WD, and by \citet{Odendaal2017}
as due to the effect of a weak ($\simeq 10^5$ G) magnetic field in an equatorial belt 
 at the boundary with an accretion disk.   

However, this model does not apply during the  thermonuclear runaway,  especially
 for the two pulsating RN/symbiotics, RS Oph and V3890 Sgr, which host
 a red giant and are thought to undergo disk accretion because
 of {\it atmospheric} Roche Lobe overflow 
 \citep[see][]{Booth2016, Miko2021}.  \cite{Ness2015} attributed the
 modulations to non-radial g-mode oscillations caused
 by the burning that induces gravity waves in
 the envelope (so called ``$\epsilon$ mechanism''),
 but detailed calculations ruled out periods longer than 
 $\simeq$10 s \cite{Wolf2018}, leaving the puzzle unsolved.

 With its excellent timing capabilities and high signal-to-noise, 
 especially in the soft X-ray range,
the Neutron Star Interior Composition Explorer Mission ({\sl NICER})
 is an excellent instrument to study the SSS and their intriguing periodic
 modulations.
The {\sl NICER} camera is an external attached payload on the International
 Space Station (ISS).  Although {\sl NICER}'s main task is
to perform a fundamental investigation of the extreme
 physics of neutron stars,  measuring their X-ray pulse
profiles in order to better constrain the neutron star equation
of state, {\sl NICER} is useful for a variety of astrophysical
 targets.  {\sl NICER} provides also unprecedented timing-spectroscopy
capability, with high throughput and low background \citep{Prigozhin2016}.
The {\sl NICER} instrument is the X-ray Timing Instrument (XTI),
 designed
to detect the soft X-ray (0.2 - 12 keV) band emission from compact sources
 with  both high time resolution and spectral information.
It is a highly modular
collection of X-ray concentrator (XRC) optics, each with an associated
 detector. The XTI
collects cosmic X-rays using grazing-incidence,
gold-coated aluminum foil optics,
equipped with 56 pairs of XRC optic modules and a silicon-drift detector for high
time resolution observations (time-tagging resolution $\leq$ 300 nanoseconds).
A shorter read-out time than other similar detectors on satellites
 in space ({\sl Chandra, XMM-Newton, Swift)} allows significantly less pile-up 
 and this is ideal for supersoft X-ray sources.

 This paper is structured as follows: each of the four source
  we studied, CAL 83, MR Vel, V1674 Her, YZ Ret,
 is described
 in a separate section, with subsections describing the data and a brief
 discussion of the results. Finally, we summarize the
 results and draw some general conclusions in a final Conclusions section.

\begin{table*}
\begin{flushleft}
\caption{Periods measured in supersoft X-ray sources.
}
\begin{tabular}{cccc}
\hline
  Object   &             Type  &            Period  &  References\\
  \hline
  RX J0648.0-4418  &    O subdwarf+WD &   13.2 s  &   \cite{Mereg2009, Popov2018} \\
                   &                  &           &   \cite{Mereg2021} \\
  N LMC 2009       &    nova          &   33.2$\pm$0.1 s & \cite{Ness2015, Orio2021} \\
  RS Oph           &    nova,symbiotic &   34.9$\pm$0.2 s & \cite{Nelson2008, Osborne2011,
 Pei2021c} \\
  KT Eri           &    nova          &   34.83$\pm$0.06 s & \cite{Ness2015, Pei2021a} \\      
  V339 Del         &    nova          &   54.1$\pm$0.2 s   &  \cite{Ness2015} \\
  CAL 83           &    main seq.+WD  &   $\simeq$67 s   & \cite{Odendaal2014, Odendaal2017} \\
  V3890 Sgr        &    nova,symbiotic &  $\simeq$82.9$\pm$0.6 s   & \cite{Beardmore2019, Page2020} \\
 Chandra r2-12    &    semi-steady SSS & 217.76$\pm$0.05 s & \cite{Trudo2008, Chiosi2014} \\  
  Nova Her 2021    &    nova          &  501.35$\pm$0.30 s &  this work  \cite[see also][]{Maccarone2021} \\
                   &                  &                    & \cite{Pei2021b, Drake2021} \\
  V407 Lup         &    nova          &  565 s &  \cite{Aydi2018} Orio2021b, in preparation \\
  XMMU J004319.4+411759 &    nova?    &  865.5$\pm$0.5 s & \cite{Osborne2001} \\
  V4743 Sgr        &    nova          & 1325 s &  \cite{Ness2003, Leibowitz2006} \\
                   &                  &        &  \cite{Zemko2016, Zemko2018} \\
  V2491 Cyg        &    nova          & 2303.4 s  &  \cite{Ness2011, Ness2015, Zemko2015} \\
  V1494 Aql        &    nova          &  2498.8 s & \cite{Drake2003} \cite{Rohrbach2009}\\
  V959 Mon         &    nova          & 55 and 102 min.  &  \cite{Peretz2016} \\
  \hline
\end{tabular}
\end{flushleft}
\end{table*}
\section{CAL 83: the ``oldest'' known SSS}
CAL 83 was discovered with {\sl Einstein} as one of the very first
 luminous SSS \citep{Long1981}.
 It is a close binary luminous SSS in the Large Magellanic Cloud 
(LMC), never observed to undergo a nova outburst. 
 The current understanding is that it hosts a massive WD accreting from a main
 sequence or slightly evolved companion, and it is luminous because
 of continuous
 shell nuclear burning. The luminosity due to the burning is orders
 of magnitudes higher than the accretion energy released in the disk.
 The accretion disk is instead a main contributor to the optical light,
 because it reprocesses X-rays from the disk in the optical energy range
\citep{Popham1996}.
 The nature of the companion is not well known. \citet{vandenheuvel1992} have
 argued that the donor star is more massive than the WD, while
 optical data analyzed by \citet{Crampton1987, Odendaal2014}
 seemed to show that the compact object is more massive than the donor.

 CAL 83 has been observed to be extremely X-ray luminous for the last
 $\simeq$40 years. Occasional ``off'' states were observed for
 at least two days in 1999 November, during one day on 2001 October,
 and again twice in 2008
 January and March, in observations spaced about
 8 weeks apart (leaving us to wonder whether
 the off state lasted for the whole 8 weeks or longer).
 A week later in  2008, detection of 
 intermediately high luminosity followed,  but after
 two more weeks - in 2008 April - the source was again undetectable
 in two exposures, with a week interval in between.  
 The sparse and irregular cadence of the observations \citep[see][]{Lanz2005,
 Odendaal2014} does not allow to draw any clear conclusion
 on the duration of the ``X-ray off'' states.  However, both \cite{Greiner2002}
 and \cite{Rajo2013} found that all ``X-ray off'' states were observed during
  optically bright states of the source, followed by a periodic
 dimming (by about 1 mag) about every 450 days. CAL 83 appeared 
 X-ray luminous when the optical luminosity was lowest. Since the
 source of optical luminosity is the accretion disk, the reason of
 this apparent anticorrelation of X-ray and optical luminosity
 may be similar to the model proposed by \citet{Southwell1996} for RXJ0513-69,
 namely that photospheric contraction follows a period
 of high mass accretion rate $\dot m$ that triggers the burning, increasing
 the effective temperature of the WD and making the WD visible as a SSS.
 A periodic decrease in $\dot m$ may be followed by quenching of the burning
 and photospheric expansion after a certain time. 

\subsection{The NICER observations of CAL83}
The details of the observations are shown in Table 2.
We obtained six observations on six different dates. 
An initial exposure of about 1 ks was done on 2019 April 19,
 to assess whether it had come out of a low state detected with Swift 
 on 2019 April 15 in a 2 ks long exposure. It was found to
 be X-ray luminous again, after little less than 4 days.
 Five more observations of very different total
 duration were done between May 16 and
 May 21 of the same year and  no more `off''
 states were observed. While the work by \citet{Odendaal2014}
 focused on XMM-Newton continuous exposures with duration of hours, the {\sl
NICER}
 observations gave the opportunity to observe if and how the period changes 
 over timescales of days and weeks.
Frequent interruptions of the {\sl NICER}
 exposures are due  to the obstruction  by the Earth, 
and the maximum exposure capacity for {\sl NICER} is limited to $\sim1000$s. 
Moreover, during effective exposures, space weather conditions can also impact 
the quality of the data when there are flares in the background,
 due to the ISS being in certain regions near the poles or the South Atlantic Anomaly.
Uninterrupted exposures of CAL 83 lasted from 208 to 1006 s, and exposures done
 on the same day, usually with intervals of
 order of $\simeq$5000 s in between, were coadded in separate observations'
 data sets with the same observation number,
 and were archived as such in HEASARC.  

\subsection{Data reduction of the NICER exposures} 
{\sc HEASoft} version 6.29c and {\sl NICER}
 data-analysis software {\sc nicerdas} version 5.0 were used to reduce the data of {\sl NICER} observations.
The timing analysis was performed with the Starlink PERIOD package,
 following the Lomb-Scargle (LS) method \citep{Scargle1982}. 
Each light curve was first detrended, by subtracting a linear fit and 
dividing by the standard deviation. The PERIOD task SCARGLE was then used to create 
Lomb-Scargle periodograms (LSP) from each light curve.
In the spectra of all the six {\sl NICER} observations of CAL 83
 all photons from the source are in the 0.2 $-$ 1.0 keV energy region, so
the light curves used to perform 
the timing analysis were extracted in this range.
With the PERIOD task
PEAKS, we found the highest peak in the periodogram between the
frequencies we specified, and in order to determine the statistical
significance of the period and obtain a statistical error, we performed
a Fisher randomization test, as described by \citet{Linnell1985}, 
over frequencies from 10 to 100 mHz, including also red
noise in the significance analysis.
We  did not subtract any background, which was very low compared
 to the source for most of the time, but we experimented with exclusions of 
 intervals of high background by using two
 slightly different methods, with the \texttt{nicer$_{-}$bkg$_{-}$estimator} tool,
 which excludes periods of inclement ``space weather'', 
and the alternative \texttt{nibackgen3C50} tool which uses a different method to 
choose the good time intervals (GTIs).
 Both tools are described in 
\url{https://heasarc.gsfc.nasa.gov/docs/nicer/tools/nicer_bkg_est_tools.html}.
However, excluding high background periods 
decreased the duration
 of the available observation and the duration
of the light curve is important for the significance of the period searching by using the LS method.

\subsection{Results of the NICER  observations}
Table 3 shows the results obtained
 with the LS period search with three differently
 filtered datasets: using all the available
 exposure time and not excluding the high background periods,
 excluding high background periods by using the 
\texttt{nicer$_{-}$bkg$_{-}$estimator} tool, and finally also 
 using the \texttt{nibackgen3C50} tool. Periodograms obtained for different dates 
 shown in Fig. 1, and the light curves folded 
with the periods listed in the second
 column of Table 3 are shown in Fig. 2. Because the actual time intervals
 considered in each exposure without exclusion of adverse space weather and
 with the two different methods are slightly different, we
 measured slightly different periods. This difference, we found,
 is actually due to the fact that the period is always
 found to drift within time scales of minutes and hours, so 
 even excluding different, short time intervals
 (that is, using different good time intervals, or ``GTIs'') 
yields different results.
 With the last correction method, the period was not detected in observation  
261100105 of May 19 2019, and there is only a low significance
 detection in the first two columns. We suggest that the reason for
 which the period becomes undetectable during short continuous exposure
 is in the variable amplitude of the modulation,
 that becomes too small for detections at times. We do not find evidence
that the periodicity is transient.  We also note that with
 both methods of excluding bad space weather, the results are consistent within
 statistical errors with  the 
``whole light curve'' in 3 out of 4 observations with a high probability detection.

 However, the detected periods are not consistent with each other
 on the different dates, showing a clear variation from one day to the
 other. The variation is not dependent on the average count rate during
 the exposure, indicating that we cannot correlate the average
 count rate with the length of the measured period.
The periodograms obtained with the dataset in column 2 of Table 3 are shown in Fig.1,
 and Fig.2 shows the light curves folded with the measured periods. 
 Table 4 shows that we did not find
 significant differences in the modulation
 amplitude of the $\sim$ 67 s period in the light curves extracted
 in both the 0.2 $-$ 0.4 keV range and in the 0.4 $-$ 1 keV range.
 Thus, the modulation is not energy-dependent in the 0.2 $-$ 1 keV in which CAL 83 is X-ray luminous.

Table 5 shows the results of the analysis of the single
 short exposures, excluding the intervals of poor space weather
 only with
 the \texttt{nicer$_{-}$bkg$_{-}$estimator} tool. Here we evaluated the
 statistical error
 for the frequency by fitting a Gaussian to the highest peak in the periodogram,
 when this fit was possible. We retrieved the frequency with high 
 significance only in 7 of the  
 single short intervals of uninterrupted exposures, and 
 the statistical error in the frequency is quite large.
 Realizing that the errors 
  are very large, we did not perform a statistical test 
 to evaluate the errors better like we did for the ``total'' observations
 in Table 3. We stress that again, 
 we did not find a correlation between average
 count rate during the exposure and frequency measured in the exposure. 
 We attribute the difficulty to measure the period during the short continuous
 observations to the varying amplitude of the pulsation. 
 Since it appears to vary when we fold the single light curves with
 the period that was detected,  we suggest in some 
 exposures it must have been too low for a clear measurement,
 and that it is unlikely that the period was transient.
\begin{table}
\begin{flushleft}
\caption{Observations of CAL 83 with NICER, exposure time,
 count rate in the 0.2-1 keV range, and ``softness ratio'' measured
 as ratio of count rate in the 0.2-0.35 keV range versus count rate 
 in the 0.35-1 keV band.}
\begin{tabular}{ccccc}
\hline
Observation ID & Start time  & Exposure time & Mean 
 count rate & Softness ratio \\
 &    (UTC) & (s)  & (cts s$^{-1}$) & (see text) \\
\hline
2611010101 & 2019-04-19,08:20:00  & 1104  & 6.000 $\pm$0.076 & 1.27 \\
2611010102 & 2019-05-16,18:35:24  &878  & 7.885$\pm$0.097 & 1.06 \\
2611010103 &2019-05-17,21:09:30  &996  & 8.197$\pm$0.093 & 1.04 \\
2611010104 & 2019-05-18,00:14:52  &11057  & 7.405$\pm$0.038 & 1.13 \\ 
2611010105 &2019-05-19,02:30:34  &2763  & 6.866$\pm$0.061 & 1.22  \\
2611010106 &2019-05-21,23:39:58  &7632  & 8.343$\pm$0.035 & 0.987 \\
\hline
\end{tabular}
\end{flushleft}
\label{tab:obscal83}
\end{table}
\begin{table*}
\begin{minipage}{155mm}
\begin{flushleft}
\caption{Timing analysis of {\sl NICER}
 light curves of CAL 83 showing the $\sim$ 67 s pulsation.}
\label{table:Temporal analysis}
\scalebox{0.76}{
\begin{tabular}{lcc|ccc|ccc}
\hline
Observation ID & Period  & Significance  & Time  & Period & Significance  & Time  & Period & Significance \\
& (s) & (\%) & (s) & (s) &  (\%) & (s) & (s) & (\%) \\
\hline
\multicolumn{3}{c|}{Whole\,light\,curve$^a$}& \multicolumn{3}{c} 
 {nicer$_{-}$bkg$_{-}$estimator$^b$} &\multicolumn{3}{c} {nibackgen3C50$^{c}$} \\
\hline
2611010101 &66.6$\pm$0.2  &99.5 & 1100  &  66.7$\pm$0.2  &99.0 & 825 & 66.0$\pm$0.2  &98.5        \\
2611010102 &65.0$\pm$1.2  & 99.0  & 877 & 65.0$\pm$1.2  & 99.0 & 759  & 65.4$\pm$1.3  & 99.5      \\
2611010104 & 65.68$\pm$0.01  &99.0   & 8435 & 66.98$\pm$0.02  &99.0 & 9388 & 66.71$\pm$0.02  &99.0     \\
2611010105 &62.42$\pm$0.08  & 44.5 & 2693 & 72.5$\pm$0.1  & 32.8 & 2010  & ...  & ...         \\
2611010106 &66.85$\pm$0.02  &99.0 &  4040  & 68.25$\pm$0.02  &99.0 & 7305 & 65.67$\pm$0.02  &99.0        \\
\hline
\end{tabular}
}
{\bf Notes:}\hspace{0.1cm} $^a $: Using light curves in which the high background periods are not excluded. $^b $: Using light curves in which the high background periods are excluded by using the nicer$_{-}$bkg$_{-}$estimator tool . $^c $: Using light curves in which the high background periods are excluded by using the nibackgen3C50 tool.\\
\end{flushleft}
\end{minipage}
\end{table*}
\begin{table*}
\begin{minipage}{155mm}
\caption{Modulation amplitudes of the $\sim$ 67 s pulsation in the 0.2 $-$ 0.4 keV and 0.4 $-$ 1.0 keV light curves of CAL 83, defined as $(max-min)/(max+min)$.  The high background periods are excluded by using the nicer$_{-}$bkg$_{-}$estimator tool.}
\label{table:Modulation amplitudes}
\begin{center}
\begin{tabular}{cccc}
\hline
      Observation& Period$^a$ & Modulation amplitude (\%) & Modulation amplitude (\%) \\
      ID&           (s) &      0.2 $-$ 0.4 keV &   0.4 $-$ 1.0 keV \\
\hline
2611010101 &66.70  & 22.4 & 27.2  \\
2611010102 &65.00  &16.2  & 17.1 \\
2611010104 & 66.98  & 4.0  & 8.2 \\
2611010105 &72.50  & 8.1  & 11.0 \\
2611010106 &68.25  &7.6  & 8.7 \\
\hline
\end{tabular}
\end{center}
{\bf Notes:}\hspace{0.1cm} $^a $: The period used to fold the 
corresponding light curves of CAL 83 in Fig.2.\\
\end{minipage}
\end{table*}
\begin{table}
\begin{flushleft}
\caption{Detailed analysis using LSP on the light curves of the single
uninterrupted short exposures of CAL 83 done with
 {\sl NICER}, corrected
 for bad space weather with the nicer$_{-}$bkg$_{-}$estimator tool.
We include observation ID (Obs. ID), observation
 number, observation segment, that is a continuous
 exposure (Seg.), starting date and time,
effective exposure duration
 in seconds (Exp.), LSP frequency with highest significance (Freq.), 1-sigma error of the LSP frequency (Err.), corresponding period (P) and significance (Sig.). Segments 8, 10, 13 in
 observation 2611010104 and segments 4, 5, 7 in 2611010106 were
 not analysed because of bad 
 space weather conditions. In the exposure column, the number
 in parenthesis is the total exposure time after the 
 exposure was corrected for bad space weather.
 No error is given for the frequency when a Gaussian fit was not possible.
}
\begin{tabular}{|c|c|c|c|c|c|c|} \hline
Obs. ID & Seg. & Start Time (UT) & Exp. (s) & Freq. (mHz) & P(s) & Sig. (\%) \\ \hline
2611010101 & 1 & 2019 April 19 08:24:36 & 405 & 14.37$\pm$0.85 &  69.6 & $<$10 \\ \hline
 & 2 & 2019 April 19 09:57:13 &           708 & 14.99$\pm$0.46 & 66.7  & 94.7 \\ \hline
 2611010102 & 1 & 2019 May 16 18:35:24  & 872 & 15.22$\pm$0.62 & 65.7  & 86.4 \\ \hline
 2611010103 & 1 & 2019 May 17 21:11:52  & 205 & --             &  --  & -- \\ \hline
 & 2 & 2019 May 17 22:44:36             & 670 & 15.22$\pm$0.85 & 65.7 & $<$10 \\ \hline
 2611010104 & 1 & 2019 May 18 00:17:19  & 769 & --             & --   & --  \\ \hline
  & 2 & 2019 May 18 01:50:01            & 791 &  --            &  --  & -- \\ \hline
  & 3 & 2019 May 18 03:24:15            & 714 & 15.26$\pm$0.65 & 65.5 & $<$10 \\ \hline
  & 4 & 2019 May 18 04:35:06            & 408 &  --            &  --  & -- \\ \hline
  & 5 & 2019 May 18 04:58:41            & 628 & 15.03$\pm$0.80 & 66.5 & 98.7 \\ \hline
  & 6 & 2019 May 18 06:07:49            & 379 & --             &  --  & -- \\ \hline
  & 7 & 2019 May 18 06:33:09            & 545 & 14.44          & 69.3 & $<$10 \\ \hline
  & 9 & 2019 May 18 08:06:37            & 519 & 14.30          & 69.9 & $<$10 \\ \hline
  & 11 & 2019 May 18 09:39:18           & 538 & 15.11          & 66.2 & $<$10 \\ \hline
 & 12 & 2019 May 18 12:41:11            & 779 &  --            &  --  & -- \\ \hline
  & 14 & 2019 May 18 14:18:26           & 507 & 15.19$\pm$0.66 & 65.8 & $<$10 \\ \hline
  & 15 & 2019 May 18 15:24:05           & 830(553) &  --       &  --  & -- \\ \hline
  & 16 & 2019 May 18 15:54:54           & 280 & 15.54$\pm$1.61 & 64.4 & $<$10 \\ \hline
  & 17 & 2019 May 18 20:02:14           & 907 & 15.02$\pm$0.63 & 66.6 & 99.9 \\ \hline
2611010105 & 1 & 2019 May 19 02:33:11   & 921 &  --            &  --  & -- \\ \hline
  & 2 & 2019 May 19 04:05:54            & 920 & 15.17$\pm$0.53 & 65.9 & 50.3 \\ \hline
  & 3 & 2019 May 19 05:39:44            & 854 & --            &  --  & -- \\ \hline
2611010106 & 1 & 2019 May 21 23:47:04   & 1006(373) & 14.86   & 67.3 & $<$10 \\ \hline
  & 2 & 2019 May 22 01:25:08            & 699 &  --            &  --  & -- \\ \hline
  & 3 & 2019 May 22 04:26:48            & 143 & --             &  --  & -- \\ \hline
  & 6 & 2019 May 22 12:21:10            & 279 & 14.78$\pm$1.86 & 67.4 & 98.9 \\ \hline
  & 8 & 2019 May 22 15:16:58            & 644 & 14.91$\pm$0.85 & 67.0 & 68.2 \\ \hline
  & 9 & 2019 May 22 16:49:09            & 854 & 15.03$\pm$0.65 & 66.5 & 99.9 \\ \hline
  & 10 & 2019 May 22 18:22:33           & 960 &  --            &  --  & -- \\ \hline
\end{tabular}
\end{flushleft}
\end{table}
%
%
\begin{figure}
\begin{center}
\plotone{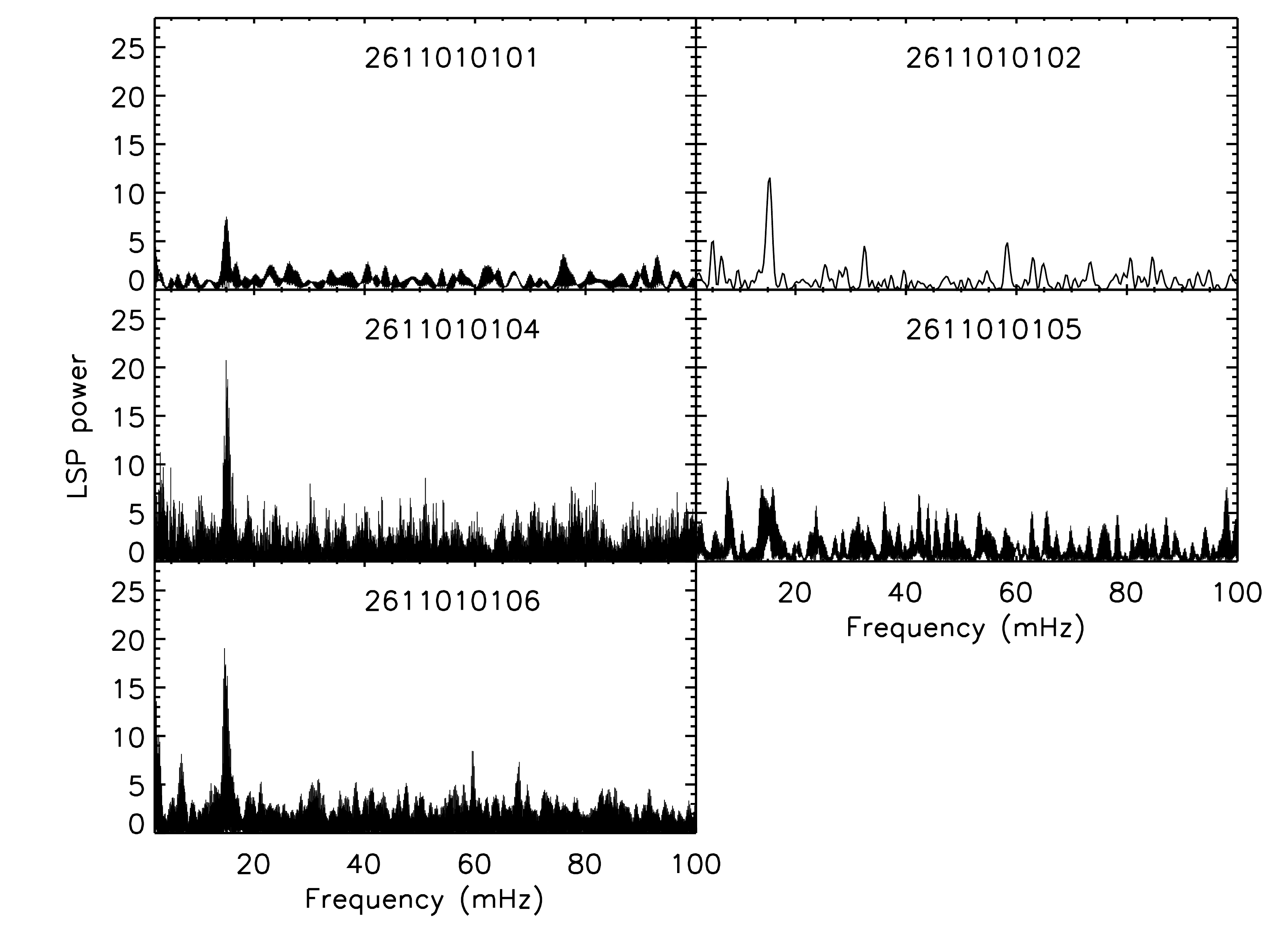}
\end{center}
\caption{Periodograms of the observations of CAL 83 done on the different dates,
 with the peaks reported in column 2 of Table 2.}
\end{figure}
\begin{figure}
\begin{center}
\plotone{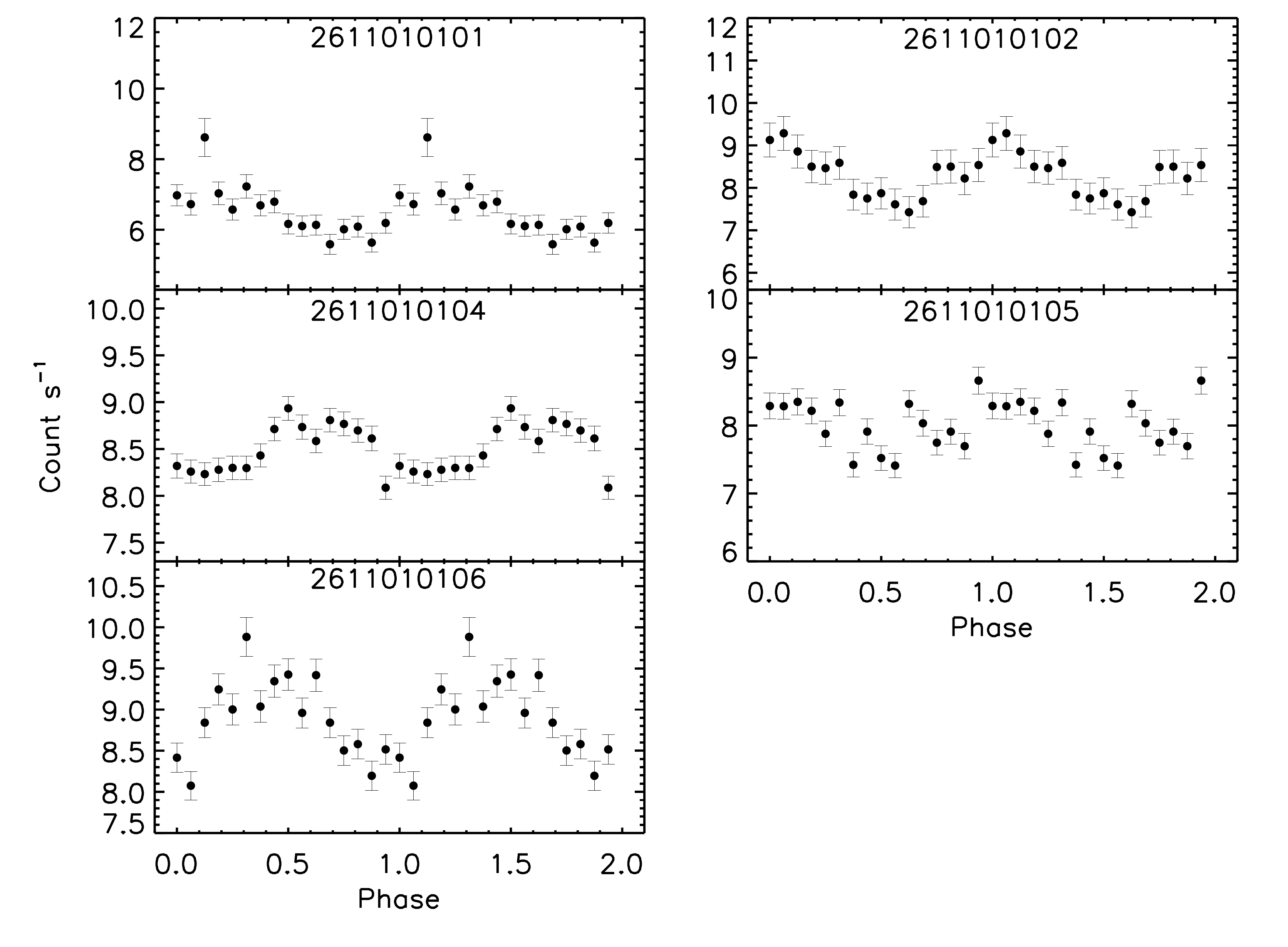}
\end{center}
\caption{Light curves of CAL 83 in the 0.2-1.0 keV range, corrected with
 the nicer$_{-}$bkg$_{-}$estimator tool, folded with the periods
 found in Table 2.}
\end{figure}
\subsection{Simulating the CAL 83 data for better understanding}
 Simulations of the periodic variability
 of nova V4743 Sgr done by \citet{Dobrotka2017} showed that a double-peak pattern 
 found in the periodogram was not caused by two actual different frequencies, 
but originated in a single frequency with amplitude
 variability, causing  a false beating. In \citet{Orio2021},
 we could not rule out that the measured
 period drift for N LMC 2009 was an artifact
 of variable amplitude, even if detailed simulations showed that an
 actual variation of the period was more likely. 
 In order to assess whether the period drift measured in CAL 83 is real, 
 or an artifact due to the variable amplitude of the oscillation,
we resorted to simulations.

The period was modeled as a sinusoidal function
\begin{equation} \label{eq:flux_xmm_all_1} \psi = \phi + a\sin(2\pi t / p) \end{equation} 
where  $\phi$ is 
 the mean period value\footnote{No long term trend was noticeable} in the observed data.
Poisson noise was then added. 
 The simulations were performed by varying either only one of the two 
 $a$ and $p$ parameters, or both of them at the same time,
 to examine 
 how the actual variations affect the amplitude or/and the period we measure.

 In order to simulate the variability, we selected randomly generated 
 points in Gaussian curves centered on
the mean value of $a$ or $p$, and selected variance values corresponding
 to typical values in the actual periodograms.
 We chose the input for the sine
 function by randomly selecting the values of a polynomial ($P_a$ and $P_p$) 
 of order between 10 and 30. If the amplitude was negative, we assumed $a = 0$.
 The sinusoidal function thus can be written as:
 \begin{equation} \label{eq:flux_xmm_all_2} \psi = \phi + P_a\sin(2\pi t / P_p). \end{equation}

 For the three types
 of variability ($var_a$, $var_a$ + $var_p$, $var_p$) we run 1000000 simulations, 
creating Lomb-Scargle periodograms. We selected the periodogram
 that best simulates the data, by calculating the sum of the residual 
 squares $\sum(o-s)^2$ calculated over a given period interval, 
 where $o$ is the observed power and $s$ the simulated one.

\begin{figure}
\begin{center}
\includegraphics[width=130mm]{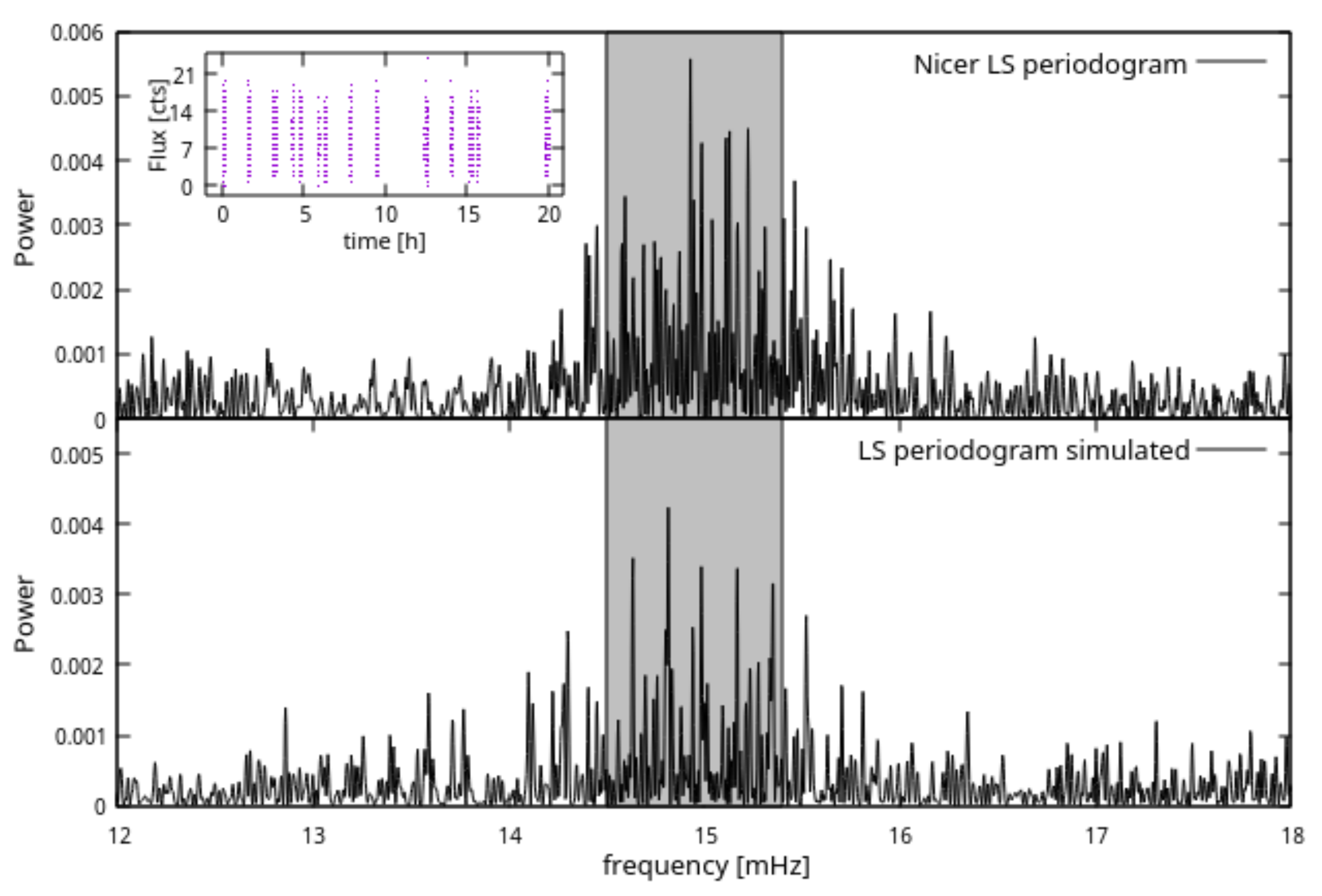}
\end{center}
\caption{The LS periodogram of the observed {\sl NICER} data of CAL 83 (top
 panel), and a simulated LS periodogram (lower panel),
 modulated with constant amplitude and constant period. 
The inset shows the corresponding light curve that consists of 15 partial observations. The gray shaded area represents the 
interval from which the $p$ values were randomly selected for the simulation.}
\label{cal83_nicer}
\end{figure}

 We performed these simulations for the light curve of {\sl NICER}
 observation 2611010104 shown in Table 5,
 consisting of 15 partial exposures (Fig \ref{cal83_nicer} inset;
 note that we neglected the intervals without any measured value, and split one 
 interval of Table 3 in two parts, because there was a small gap during
 the exposure).
 Besides the detected $\sim$67 s period, some randomly selected $p$ mean values between 
 65 s and  69 s (gray shaded area in Fig \ref{cal83_nicer}) were
 also  used. We chose the best 100 results for each type of variability
 by examining the sum of the residual squares, and compared the observed and 
simulated LS periodograms. 
52 of the
 best periodograms were simulated with variable amplitude and constant period, 23 with
constant amplitude and variable period and, 25 with variable amplitude and period. 

The result indicates that the constant and variable period fit
 the data equally well.  However,
Fig \ref{cal83_nicer} (bottom panel) shows the simulated LS periodogram in which the 
light curve was modulated with constant amplitude and constant period. It
 is clear that it is affected by significant aliasing because of the 
 different duration of the partial observations and intervals 
 between them. Thus, complicated patterns of variability cannot be analysed.

In order to eliminate the aliasing problem, we repeated the simulation 
with the same $p$ values, splitting the light curve in the short
 continuous exposures listed in Table3, 
and created an LS periodogram for each short exposure. 
The duration of the individual continuous
 exposures varies from 240 s to 907 s 
 and, as shown in Table 3, was often not sufficiently long
 for  clear  detections of the periodicity. 
In the simulations, we found the following distribution 
 for the 100 periodograms with the best  residuals:
 48 with variable amplitude and constant frequency, 17 with constant
 amplitude and variable frequency, 35 with variable amplitude and 
variable frequency.  However, the LS periodograms have very broad peaks,
 which in most cases we could not reproduce, 
 so this simulation did not lead to a clear conclusion.
Fig. 2 of \citet{Odendaal2014}, based on  {\sl XMM-Newton} data, shows
 that the width of the frequency drift 
due to potentially variable periodicity 
is approximately 1-2 mHz (see ObsID 0506531701). Considering
 1/T (where T is the duration of a continuous exposure) 
as periodogram resolution, even the longest {\sl NICER}
snapshot of 907 s yields a resolution of 1.1 mHz, which is 
comparable to the width of the frequency drift and does not allow 
 a detailed analysis. As a test, we repeated the simulations only with
 exposures longer than 500s (a limit that
 was empirically based on visual inspection of the periodograms) and found
 a different result, namely 94 of the 100 best simulations had
 variable amplitude and fixed frequency. Thus, 
the results of
 these simulations of the many short {\sl NICER} exposures depend
  on the minimum length chosen to select the exposures to model. 

\begin{figure}
\begin{center}
\includegraphics[width=130mm]{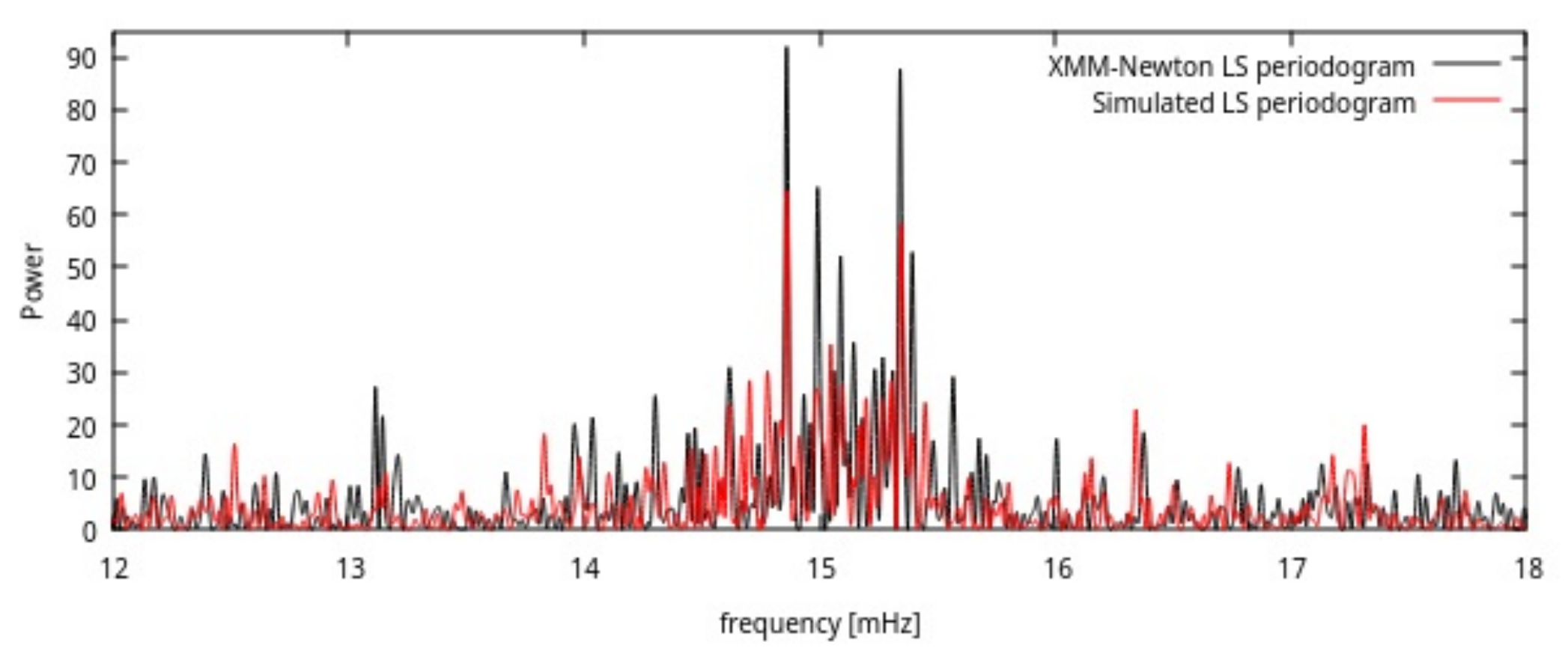}
\end{center}
\caption{The best simulated LS periodogram for the XMM-Newton observation
 of CAL83 started on 2009-05-30 and lasting 46 ks.
 The light curve was modulated with constant amplitude and variable period.}
\label{cal83_xmm}
\end{figure}

We conclude that the analysis should be based on much
 longer continuous exposures and repeated the simulation for the \textit{XMM-Newton} EPIC-pn light curve 
of ObsID 0506531701, the most recent
 and longest continuous exposure among all
 X-ray observations of CAL 83  \citep[see][]{Odendaal2014}. 
As Fig. 3, bottom right panel, shows, this continuous exposure gave
 a complex result, with other minor peaks in the periodograms that
 were relatively high.  We adopted the 
 following mean values: $p$ = 65.2s ,  $p$ = 66.3 s,  $p$ = 66.7 s, $p$ = 67.3 s,
 corresponding to the highest peaks around the observed $\sim$67 s periodicity, and a value
 between them. We modeled the long-term trend of the observed light curve by
 assuming that the vertical shift $\phi$ resulted from 
a polynomial fit (13th order, $P_{13}$). Again,
 we selected the 100 best residuals fits, obtaining
 the following outcome: 87 were with constant amplitude and variable period, 13 with
 variable amplitude and variable period, none with a constant period. 
The best simulated LS periodogram is shown in Fig \ref{cal83_xmm}.\par

 With the result obtained with the XMM-Newton data 
we draw the conclusions that CAL 83 has
 a variable period,  ruling out that the observed LS periodogram is
 modulated by a stable period with  variable amplitude (even
 if also the amplitude varies).  As mentioned above, 
\cite{Dobrotka2021} were able to model the periodogram of Nova LMC 2009a in
 the SSS phase even only with variable amplitude, but a variable period appeared
  to yield better simulations. 
 While in the LMC 2009a the scattering was around 0.4 s, in CAL 83 it is 
 quite longer, namely 2.1 s,  allowing a more definite conclusion.
\subsection{CAL 83 spectra observed with NICER}
CAL 83 was observed with the {\sl Chandra} Low Energy Transmission Grating 
(LETG) in 2001 May and with XMM-Newton once in 2001
 and in several exposures between 2007 and 2008. Two LETG
 exposures in 2002 October found the source in an ``off'' state.
  Because of the shorter {\sl NICER} read-out time, 
 the spectra are not affected by pile-up in this extremely soft source,
 like EPIC and the Chandra ACIS-S (an observation
 was done with this instrument in 1999), but CAL 83 is too soft for the
 {\sl XMM-Newton} RGS grating, that does not measure the whole spectral
 range of interest. Therefore, we compare the {\sl NICER} spectra with the 
 ones observed in
 2001 with the LETG, despite the very different spectral resolution. 
 The average count rate of CAL 83 has always varied in different
 observations \citep[see][]{Odendaal2014}, and an interesting question
 is whether there is a correlation with spectral variations.
 In Fig. 5, in the top panel on the right, we compare 
 the {\sl NICER} average spectra on different days and we also show the
 comparison with the LETG spectrum of 2001. 
 One of the 2019 NICER spectra, and the LETG 2001 one, are
 shown fitted with atmospheric models in the lower panels of Fig. 5, in
 units of energy and wavelength, respectively.
 A rigorous fit was done by \citet{Lanz2005}, who
 found a best fit with  with log(g)=8.5 and 
 T$_{\rm eff}$=550,000$\pm$2,500 K,
 fixed N(H)=6.5 $\times$ 10$^{20}$ cm$^{-2}$, absolute
 luminosity 1.8$\pm$0.6  $\times$ 10$^{37}$ erg s$^{-1}$.
 These author's atmospheric models are not public, so we used the publicly available TMAP code
 grid of models of \citet{Rauch2010}, available in the web site
 \url{http://astro.uni-tuebingen.de/#rauch/TMAP/TMAP.html}.
 We find the best fit to the {\sl NICER} spectrum of May 18 (the longest
exposure) with  
 a model studied for Magellanic Cloud and Galactic halo sources
 with depleted abundances, with log(g)=9.0. 
 Model ``halo'' with log(g)=9, that we used, is adopted from a grid of models
 with only elements from H to Ca, that is defined as ``not suitable for precise spectral analysis''
 because only approximate formulae were used in order to account for
 Stark line broadening. However, with 
 the broad-band spectra the absorption
 lines are not measurable and great precision is not required.
 The more rigorously calculated grid with elements
 up to Ni has peculiar abundances adopted from specific
 nova models \citep[see][]{Rauch2010}, so we found that the depleted 
abundances model is more suitable, as expected because 
 the Magellanic Clouds  have low
 metallicity and no strong mixing like in post-novae should
 have occurred in CAL 83.
 All the public
 grids in the web site are either with log(g)=8 or log(g)=9, but  
 no available models were computed with values between these two extremes. 
 log(g)=9 models fit better than those with log(g)=8, but
 the fit we obtain is not perfect, mainly because it underpredicts
 the flux above 0.5 keV. The values we obtain
 are close to those of \citet{Lanz2005}, but the higher
 log(g) results in a lower temperature, namely
 T$_{\rm eff}$=502,000$\pm$5,000 K, N(H)=5.1 $\times$ 10$^{20}$ cm$^{-2}$, 
 absorbed flux 7.4 $\times$ 10$^{-12}$ erg s$^{-1}$ cm$^{-2}$
 and unabsorbed flux 3.5 $\times$ 10$^{-11}$ erg s$^{-1}$ cm$^{-2}$.
 We find that this model fits also the LETG spectrum of 2001,
 with T$_{\rm eff}$=469,000$\pm$5,000 K, N(H)=6.3 $\times$ 10$^{20}$ cm$^{-2}$, 
 absorbed flux 8.3 $\times$ 10$^{-12}$ erg s$^{-1}$ cm$^{-2}$,
 unabsorbed flux 5.6  $\times$ 10$^{-11}$ erg s$^{-1}$ cm$^{-2}$
 (corresponding to an absolute
 luminosity of 1.67 $\times$ 10$^{37}$ erg s$^{-1}$ at a distance
 of 50 kpc to the LMC).  However, 
 if we model the LETG spectrum with more bins, we obtain a better result with 
 the grid studied for nova abundances, 
 (specifically, model SSS-003-00010-00060.bin-0.002-9),
 because it includes more ions and thus models the absorption features.
 The resulting parameters are
 T$_{\rm eff}$=505,000$\pm$5,000 K, N(H)=5.8 $\times$ 10$^{20}$ cm$^{-2}$,
 absorbed flux 7.9 $\times$ 10$^{-12}$ erg s$^{-1}$ cm$^{-2}$, 
  unabsorbed flux 4.5  $\times$ 10$^{-11}$ erg s$^{-1}$ cm$^{-2}$ 
(corresponding to absolute
 luminosity 1.34 $\times$ 10$^{37}$ erg s$^{-1}$).
 Since we did not have a sufficiently ``sensitive'' grid of log(g) values,
 the small differences in the best fits' parameters
 (of order 10\%) are within each other's statistical
 errors and do not indicate any significant spectral difference between 2001 and 2019.
\begin{figure}
\begin{center}
\includegraphics[width=83mm]{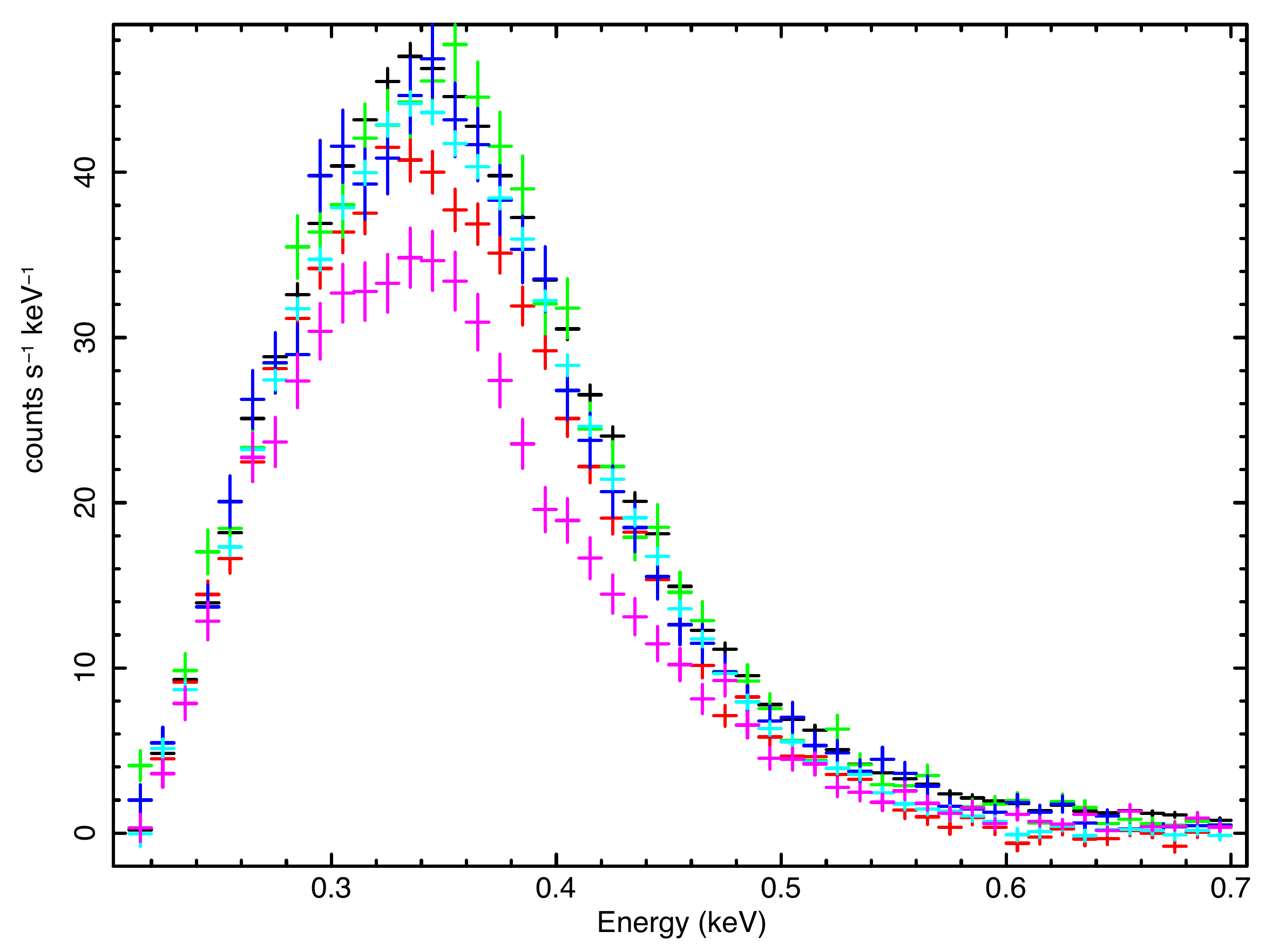}
\includegraphics[width=83mm]{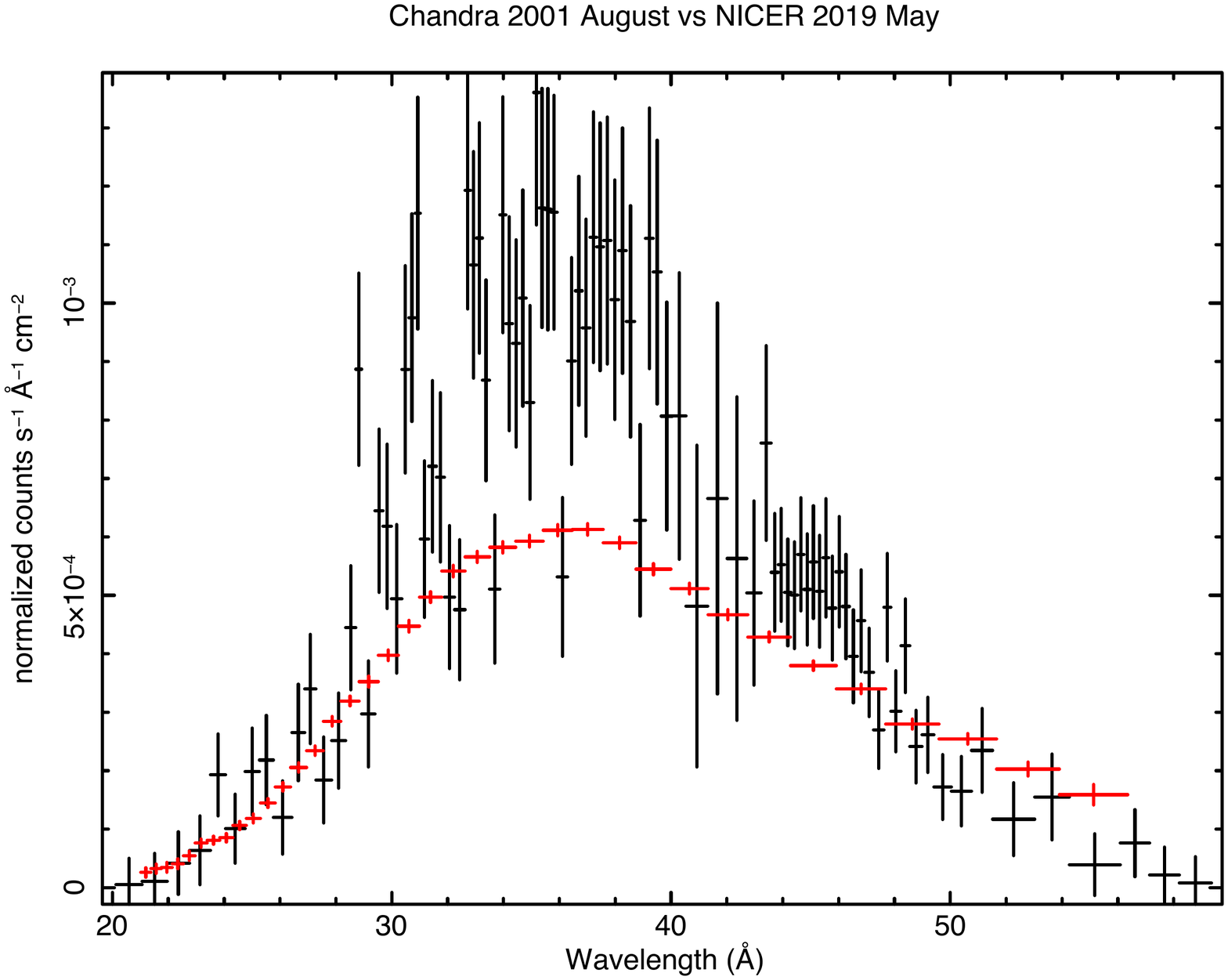}
\includegraphics[width=83mm]{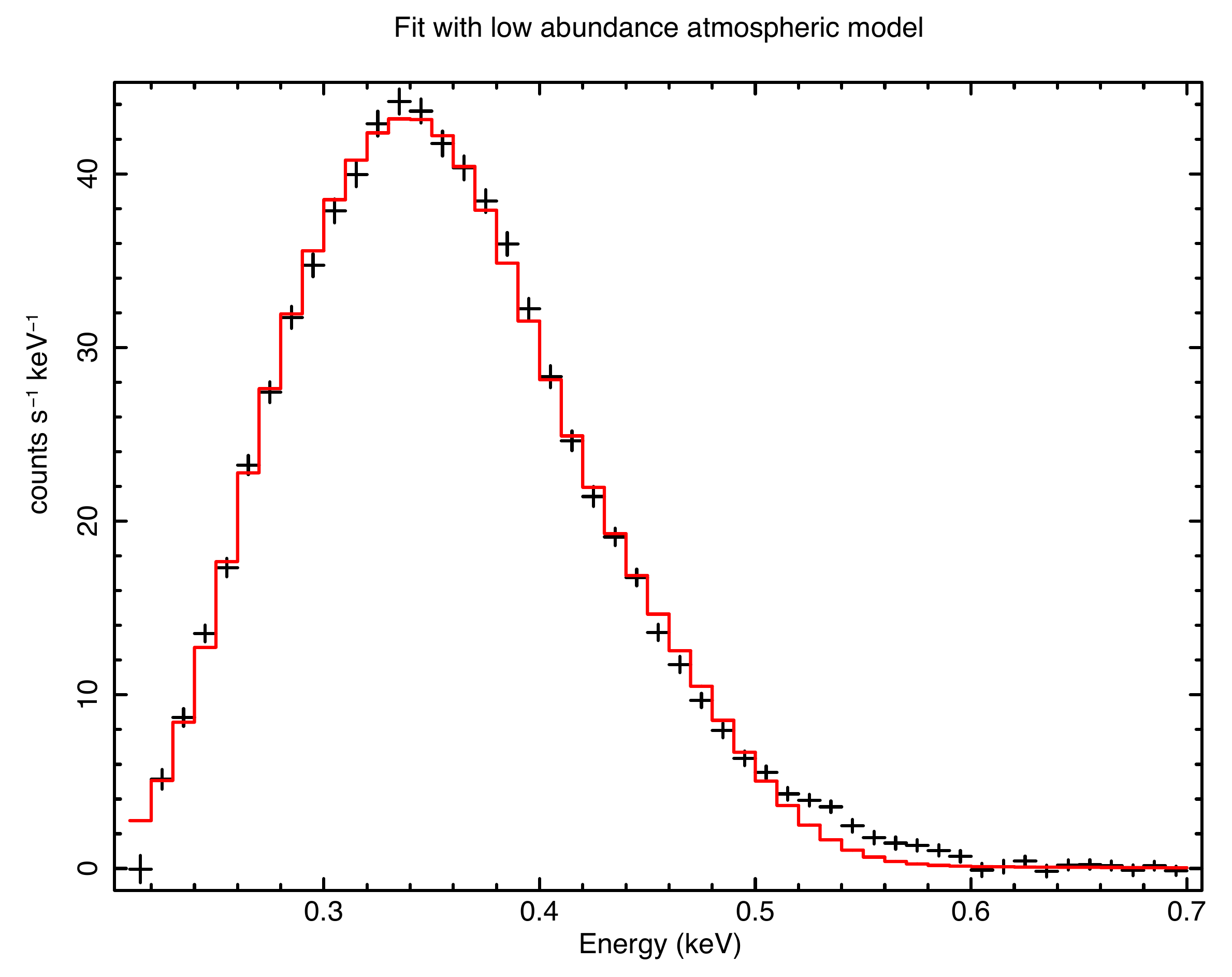}
\includegraphics[width=83mm]{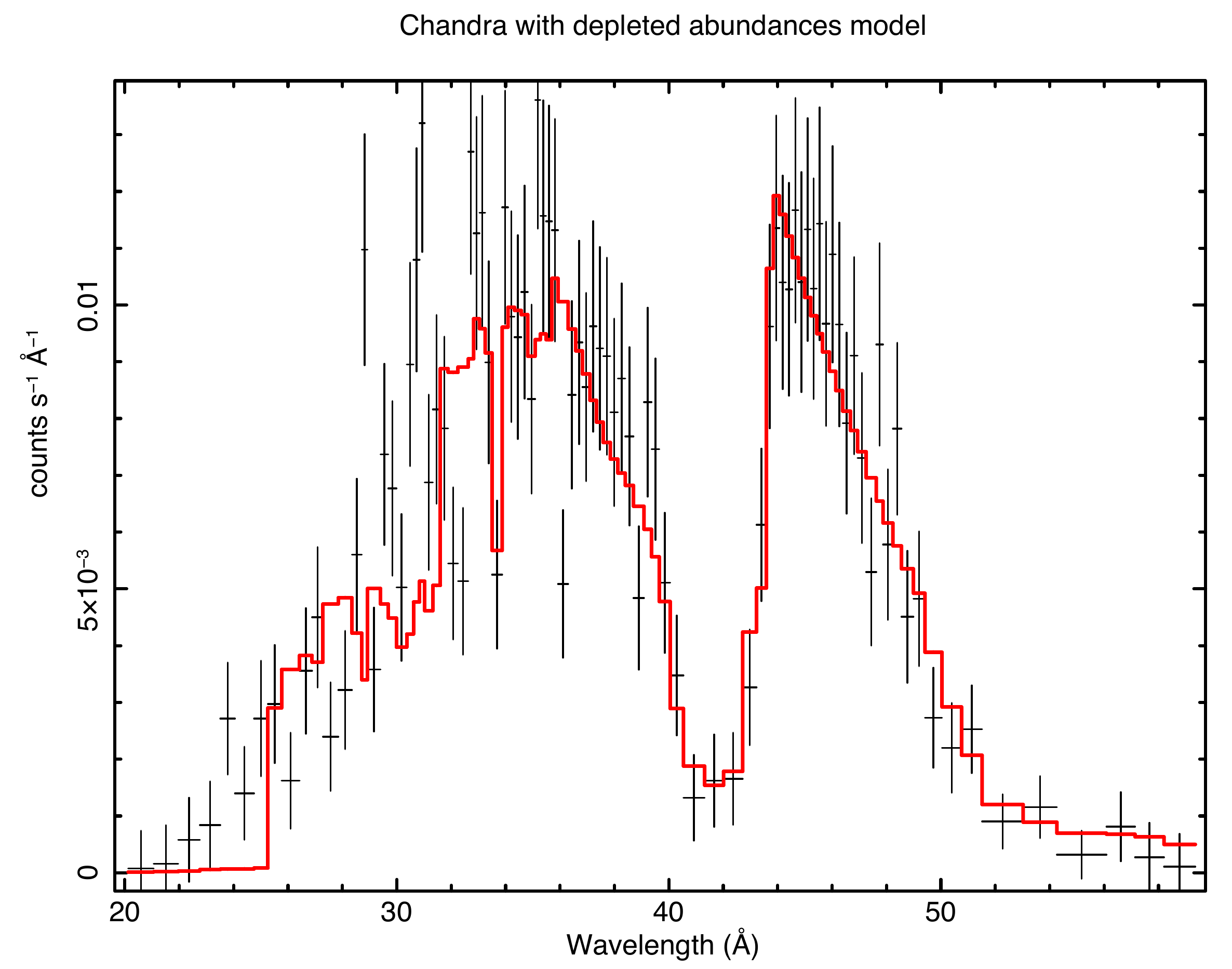}
\end{center}
\caption{The top panel on the left shows the CAL83 average 
count rate spectra measured during 
 each day of {\sl NICER} observations, done in 2019. 
 The top panel on the right, in units of
 wavelength appropriate for the grating spectrum,
 shows the comparison between the fluxed spectrum  (normalized for the effective area)
 measured on May 21 with {\sl NICER} (the one with the largest count rate)
 with the fluxed
 2001 Chandra LETG spectrum (again as a function of
 wavelength to respect the gratings scale),
 binned with large bins of signal
 to noise - at least S/N=100 per bin - for a better comparison
 with {\sl NICER}. There was excess flux in 2001, but
 the peak and spectral shape seem unchanged. In the bottom panels we show 
 fits to the 2001 Chandra (right) and 2019 {\sl NICER} (left) 
count rate spectra,  respectively, with the abundance depleted
 atmospheric model discussed and referenced in the text and the same parameters.
}
\end{figure}

Because the peak of the count rate
 is around 0.35 keV, in Table 2 we also give
 the ``softness ratio'' measured as count rate in the 0.2-0.35 keV
 band divided by count rate in the 0.35-1 keV band. It is
 evident from the table that this ratio decreases when the total
 count rate increases, thus when the average count rate 
 is lower, the softest portion increases. We also tried to estimate
 the flux in two energy ranges, the softest, namely 0.2-0.3 keV and the total
 flux in the 0.2-0.8 keV range (there is hardly any
 flux above 0.8 keV).  
 We did so by fitting the model described above, choosing on purpose
 the softest possible range for this comparison. We found that   
 the unabsorbed flux in this softest band varies by at most
 15\% around an average $\simeq$ 2.1 $ \times 10^{-12}$
 erg cm$^{-2}$ s$^{-1}$. The total range in the whole 0.2-0.8 keV range
 of interest also varies by about 15\% 
 around an average value of $\simeq 7 \times 10^{-12}$
 erg cm$^{-2}$ s$^{-1}$.  From these two indications, ``softness
 ratio'' and comparison of flux estimates, 
 we infer that the measured count rate variation
 is not due to change in column density N(H), as would be the case 
 if it was due, for instance, to a wind.

\subsection{Accretion and burning in CAL 83}
 \citet{Odendaal2017} offered an attractive explanation for the oscillations in the
 X-ray flux of CAL 83, as a ``dwarf nova oscillation'' type of quasi periodic oscillation in
 a ``belt-like structure'' at the boundary of the inner accretion disk, which is
 weakly coupled with the WD core by a $\simeq 10^5$ G magnetic field, rotating faster
 than the WD itself and with a combination of spin-up and spin-down. However, the
 high S/N obtained with {\sl NICER} in the softest range (0.2-0.4 keV) demonstrates
 that the modulation is not energy dependent.  
 The accretion belt is  very unlikely to emit such conspicuous flux
 in the very soft energy band of 0.2-0.4 keV. Since we attribute the 
 extremely high flux in this range only to the WD atmospheric emission,
 the pulsation appears to be occurring on the WD.
  
 \citet{Odendaal2017} did not quantify all aspects of the model for
 the case of  CAL 83 and spectral predictions
 are missing in the paper, but since
 accretion would produce a much less luminous and much
 less soft emission than observed, the absence of energy dependence 
may be a strong
 point against the so called eLIMA model of these authors.
 The short period oscillations with a period drift observed 
 in RS Oph and V3890 Sgr,
 which are symbiotic systems, during their nova eruption are also
unlikely to be explained by such a model. The
 accretion disks in symbiotics have very different size and characteristics 
 and there is evidence that
 at least the inner accretion disk is destroyed during the eruption
 \citep{Worters2007}.  In the 2006 outburst the accretion
 disk of RS Oph was only re-established by day 241 of the outburst,
 long after the SSS turn-off \citep{Worters2007}. A common model for all the shortest period modulations of
 the SSS seems desirable, even if only for an ``Occam's razor'' criterion.

   Another important point raised by the {\sl NICER} monitoring is that the SSS flux
 always appears to vary by up to 30\%, even within hours, without spectral
 hardening that may indicate increased absorption. Moreover, we observed
 that the rise from the minimum X-ray luminosity 
 L$_{\rm X} \leq 10^{36}$ erg s$^{-1}$ to an X-ray
 luminosity level above 10$^{37}$ erg s$^{-1}$ took only 3 days, and the source was never
 observed again at minimum in the course of 5 days in 2019 May. A successful explanation
 of the luminosity variations of CAL 83, attributed either to photospheric
 adjustments or to variable accretion \citep[see][]{Greiner2002}, has to take these
 timescales into account.  
\section{MR Vel: an ``extreme'' symbiotic star}
 MR Vel (RX J0927.5-4758) is a peculiar steady X-ray source,
 initially observed with {\sl ROSAT}
 and later observed and found equally luminous with {\sl ASCA, BeppoSAX,
 Chandra and XMM-Newton},  with a highly absorbed
 (N(H)$\simeq 10^{22}$ cm$^{-2}$), yet supersoft
 X-ray spectrum due mainly to emission lines \citep[see][and references
 therein]{Motch2002, Bearda2002}. Despite the relatively low
 inclination, estimated to be between  $50\degr \leq i \leq     
60\degr$ \citep[according to][]{Schmidtke2000},
 the X-ray grating spectra show an emission line spectrum that cannot
 be fitted with models of collisional ionization or photoionization 
 and is difficult to explain.
  The absorbed X-ray luminosity is about 10$^{35}$ erg s$^{-1}$ at a distance
 of 1 kpc, but GAIA indicates a value of 5.68$_{-0.74}^{+1.00}$ kpc
 \citep{Bailer2021}.

 The orbital period of MR Vel is 4.028 days
 \citep{Schmidtke2000, Schmidtke2001}, and
 the optical light curve also shows modulations with
 periods of 0.2 to 0.3 days - a 0.2 days modulation was  also observed in 
 X-rays in the year 2000 \citep[see][and references therein]{Motch2002, Bearda2002}.
 The radial velocities of the HeII 4686 and H$\alpha$ emission lines
and their amplitude allowed to estimate a mass range of 1-2
M$_\odot$ for the donor, and of 0.7-1.7 M$_\odot$  for the accreting
 object \citep{Schmidtke2000, Schmidtke2001}. 
 So far, no trace of the donor has been detected in the optical
spectrum \citep{Schmidtke2000}, nor in the I-band \citep{Mennickent2003}.
In June 1997, \citet{Motch1998} detected transient jets from MR Vel with
projected velocity of 5200 km/s.  The receding component of the
H$\alpha$ jet with projected velocity of 5350 km/s was tentatively
identified by \citet{Schmidtke2000} in March 1999. Since for the majority of
jets, the outflow velocity reflects the escape velocity at their origin
\citep{Livio1998}, and assuming that the jet expansion
is perpendicular to the orbital plane, the expansion velocity is $\sim 8100
\leq V_{\rm exp}=V_{\rm jet} {\cos i}^{-1} \leq 13000$ km s$^{-1}$ 
for the inclination range given above ($50\degr \leq i \leq 60\degr$), is
 consistent with a massive, $\sim 1-1.3$ M$_\odot$, WD
accretor  and narrows the range for the
mass of the donor, $M_1 \sim 1-1.6$ M$_\odot$.
For a binary period of about 4 days and the above mass range, the Roche
lobe radius of a main sequence secondary should be of the order of $R_2 \sim 1-1.5$
R$_\odot$, thus a donor star with mass $\sim$ 1-1.6 M$_\odot$ fitting
 this system must be
evolved. The effective temperature of this donor does not exceed $\sim$ 6000
K, and the luminosity does not exceed $\sim 30$ L$_\odot$.
The 2MASS magnitudes of MR Vel, $J=12.934 \pm 0.029$, $H=12.204 \pm
0.102$, $K=11.770 \pm 0.024$ provide some additional constraints. 
The source is highly reddened, as indicated by the neutral
hydrogen column density, $N_H \simeq 10^{22}$ cm$^{-2}$, 
which corresponds to a reddening of E(B-V)$\simeq 1.4 - 1.6$ 
\citep[adopting the conversion
relationship given by][]{Guver2009}. The
reddening corrected $(J-K)_0 \sim 0.4-0.5$ is thus consistent with a G2-8
giant ($T_{eff} \sim 5000 - 5600$ K; moreover  the magnitude $K_0 \sim
11.2 - 11.3$ combined with the Gaia distance indicates a luminosity of
$\sim 110 -180$ L$_\odot$.  The radius of such  a star, $\sim
14$ R$_\odot$, would be almost 3$\times$ larger than the Roche lobe radius.
Summarizing, the observational data are consistent with a low-mass,
1-1.6 M$_\odot$, yellow (G-type) giant donor ascending the first
giant branch, and a massive, 1-1.3 M$_\odot$, WD secondary.
So, in principle, MR Vel is likely to be a sort of  extreme (with the
shortest orbital period!)  yellow symbiotic system, similar to, e.g.,
StHA 190 \citep{Smith2001}.  
\begin{figure}
\begin{center}
\includegraphics[width=140mm]{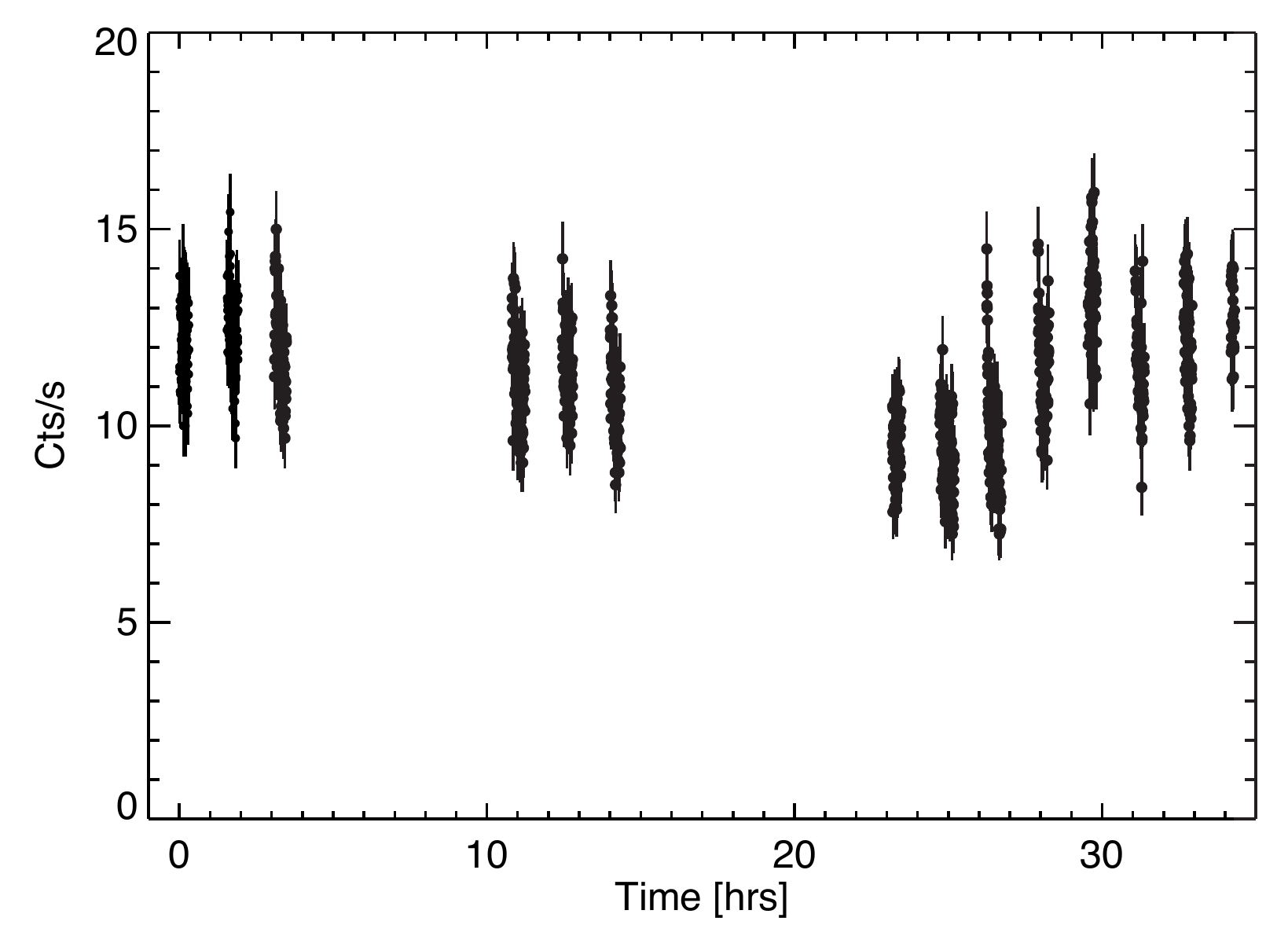}
\end{center}
\caption{The lightcurve of MR Vel
observed with NICER from May 19 2019 until May 20 2019
with time bins of 16 s.}
\end{figure}
\begin{figure}
\begin{center}
\includegraphics[width=83mm]{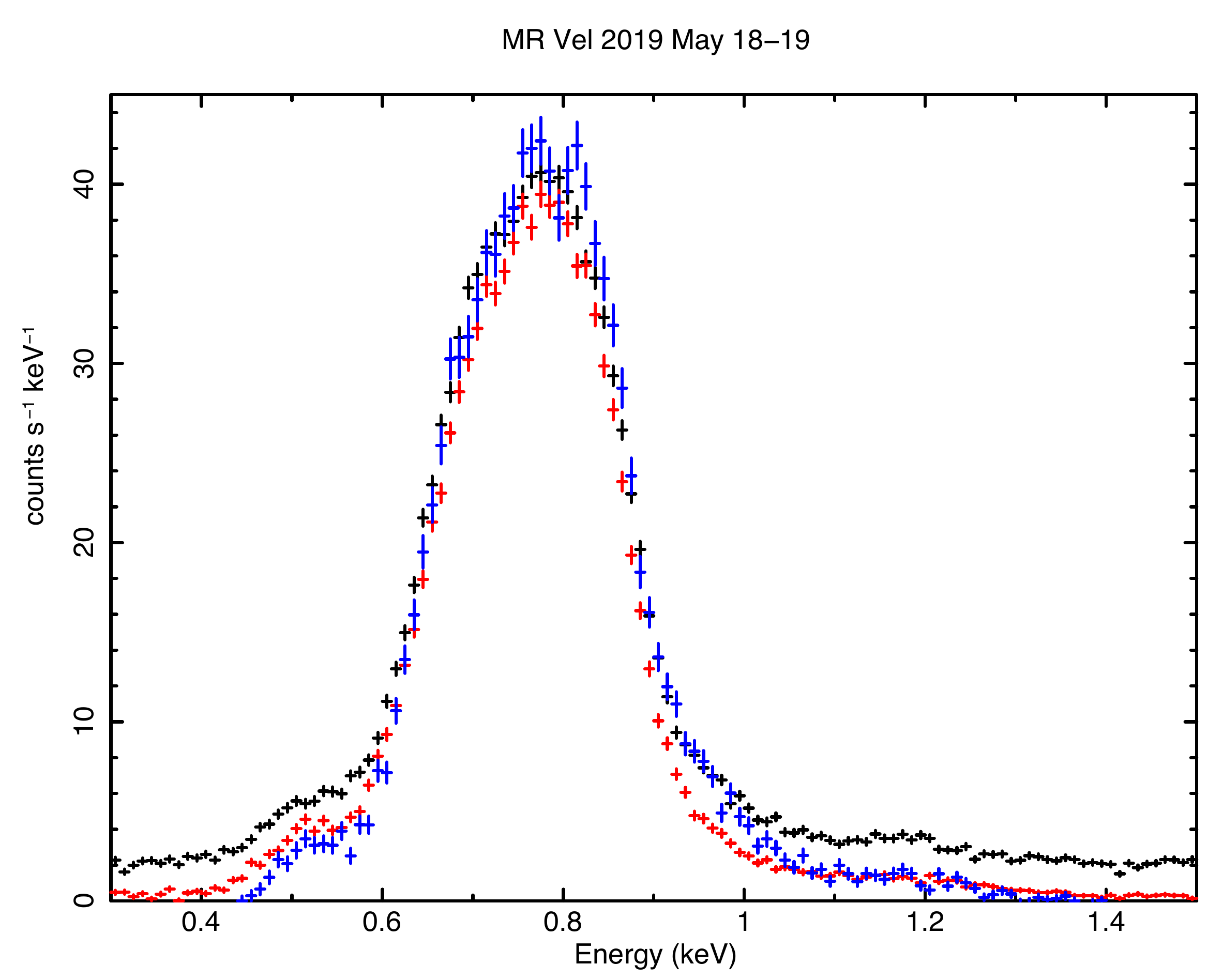}
\includegraphics[width=83mm]{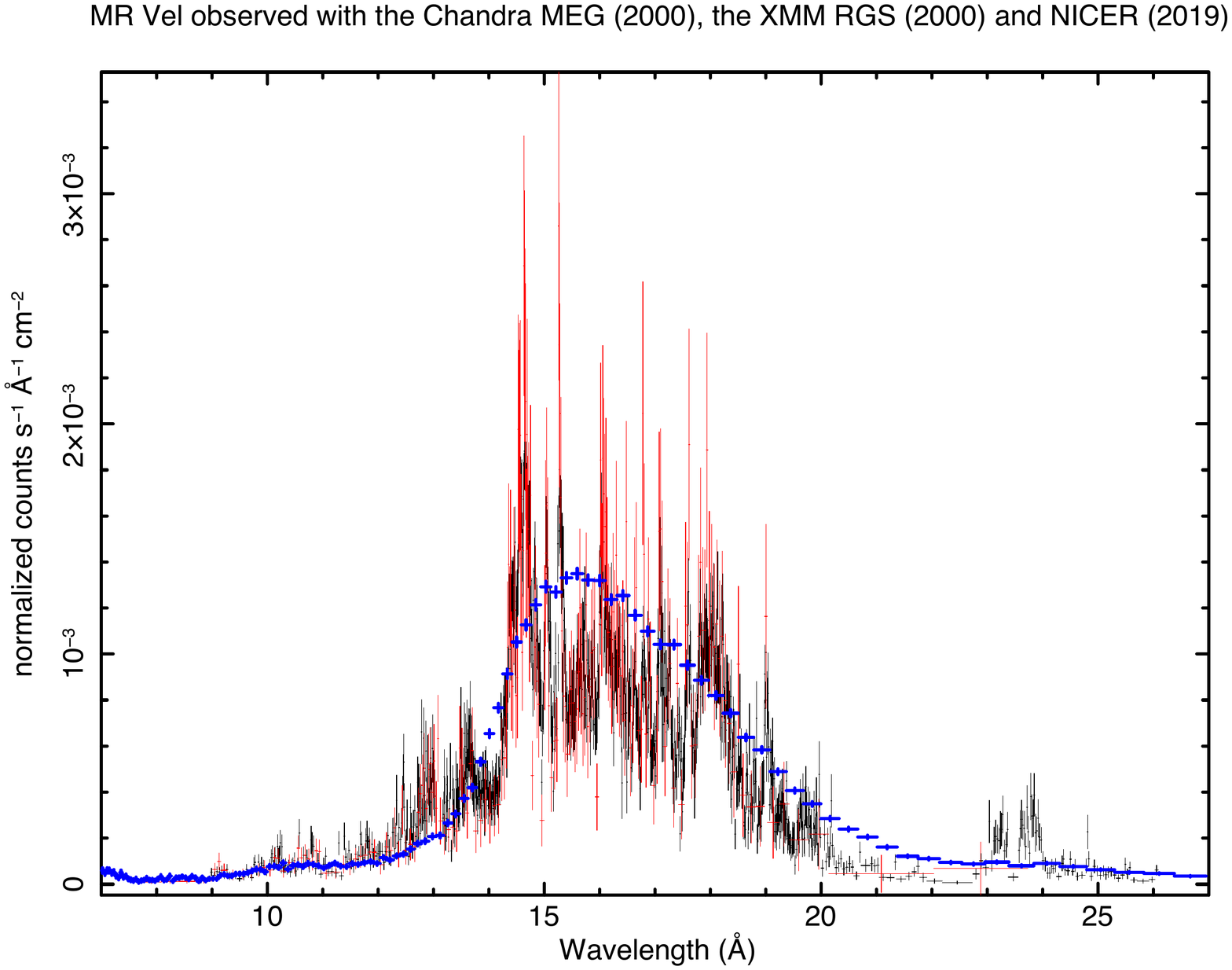}
\end{center}
\caption{On the left, the energy/count rate
 spectra of MR Vel observed with {\sl NICER} in three different
 observations composed of coadded individual exposures (see Fig. 6)
 and on the right, comparison of the grating
 spectra, fluxed (normalized for effective area) and plotted in their natural wavelength units,
 observed in 2000 with the Chandra MEG grating
 (red), and with the XMM RGS gratings (black) with
 the  2021 May 18 observed with {\sl NICER} (blue,
 plotted here also as a function of wavelength for comparison.)}
\end{figure}
\begin{figure}
\begin{center}
\includegraphics[width=150mm]{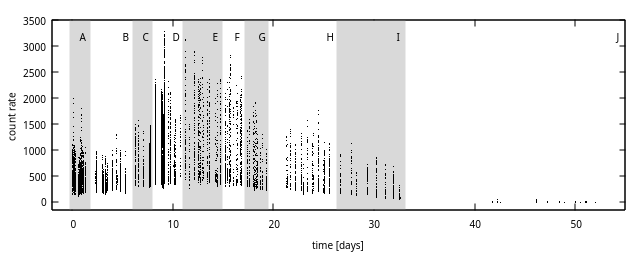}
\end{center}
\caption{The light curve of V1674 Her observed with NICER since 2021
 July 10 at UT 18:28:40, binned with 16 s
 bins. The very large modulation amplitude with the $\simeq$501
 s period in each single
 observation, and superimposed random variability,
 is evident from the spread of the points.}
\end{figure}
\begin{figure}
\begin{center}
\includegraphics[width=175mm]{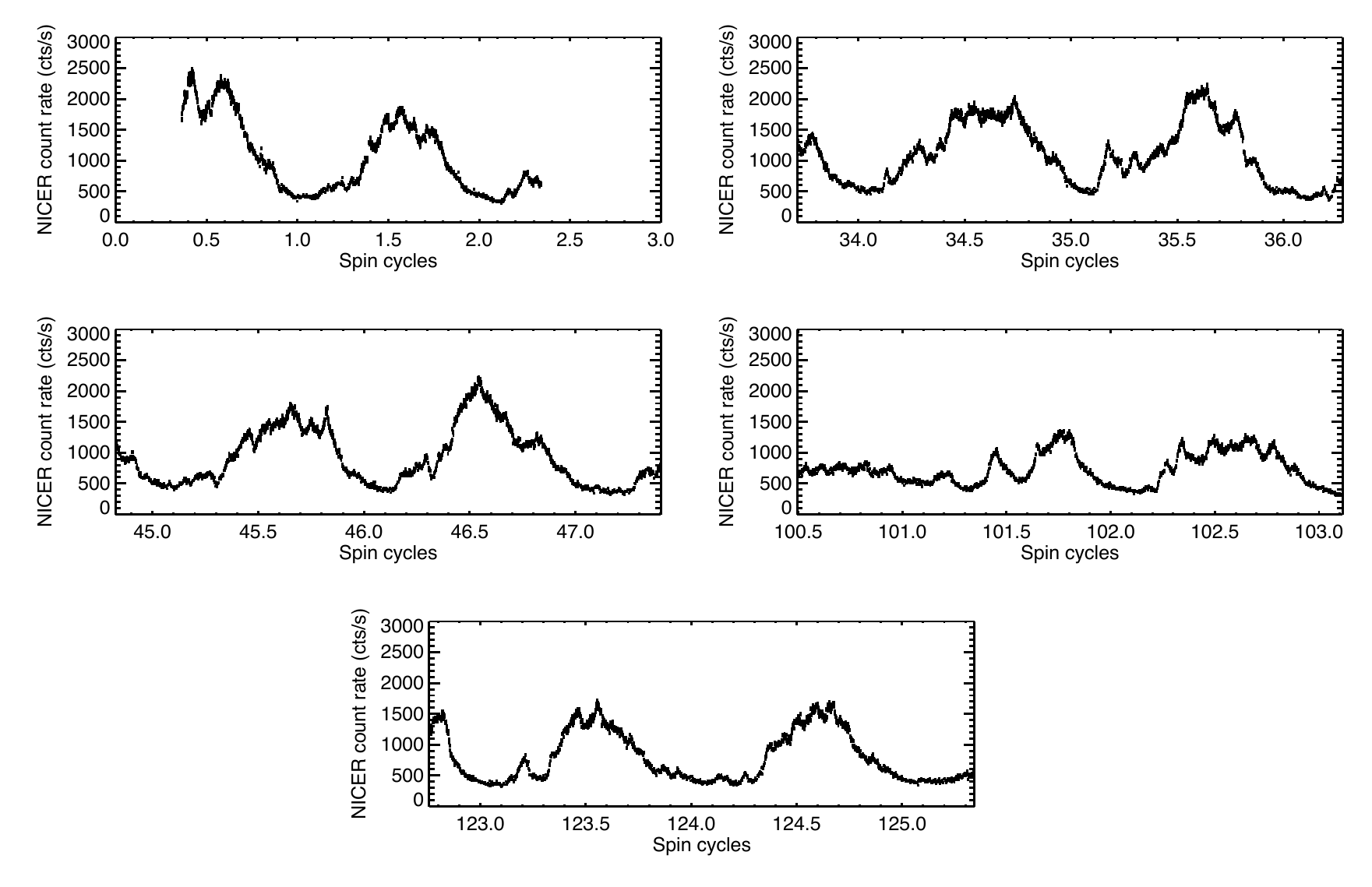}
\end{center}
\caption{Light curves of the GTIs of July 20 observations of V1674 Her 
(4202260108),
 during interval D of Fig. 8, binned with 1 s bins and plotted as a function of 
 the elapsed phases, assuming a zero point in the first minimum observed
 on that day.}
\end{figure}
\begin{figure}
\begin{center}
\includegraphics[width=160mm]{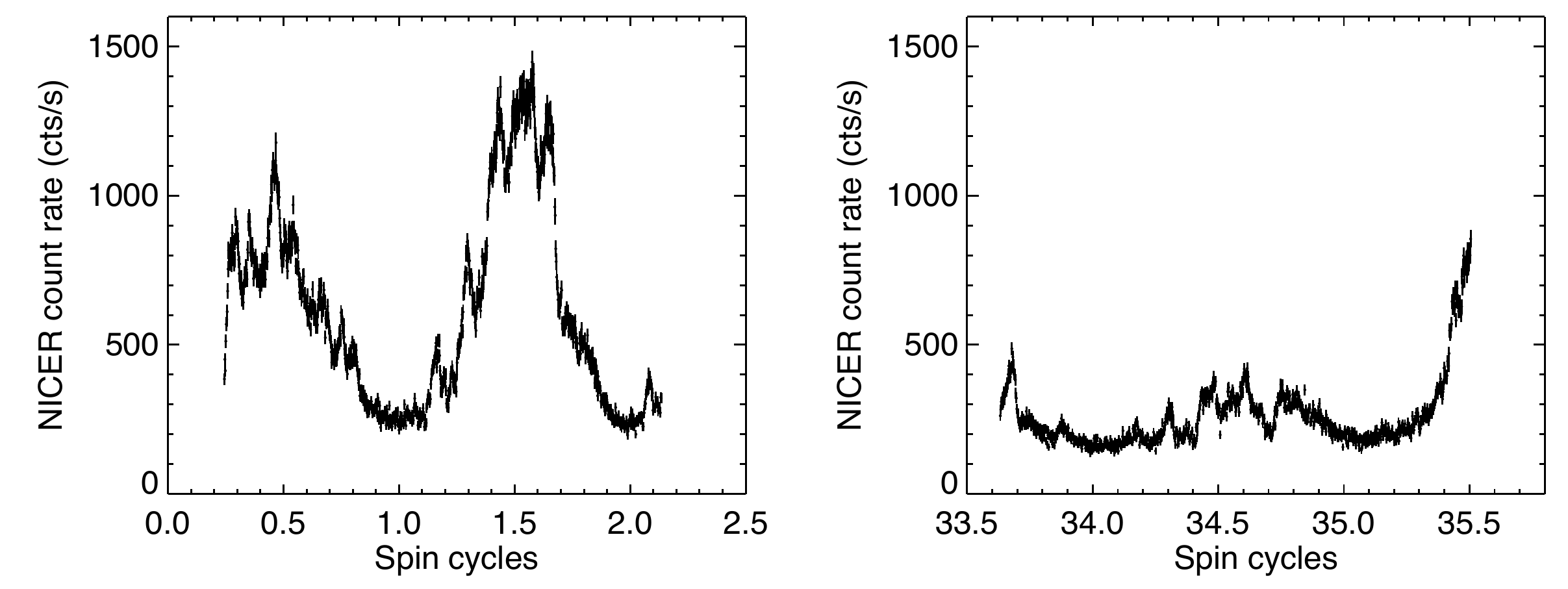}
\end{center}
\caption{Light curves of the GTIs of August 2  observations of V1674 Her
(4202260120),
 during interval G of Fig. 8, binned with 1 s bins and plotted as a function of phase, again assuming a zero point in the first minimum of that day.}
\end{figure}
\subsection{The NICER observations of MR Vel}
 We observed MR Vel with {\sl NICER}  in several exposures lasting 
 about 1000 s each on 2019 May 18 and 19. 
 The data reduction was performed as described above for CAL 83,
 and the data we present had bad space weather intervals
 subtracted with the \texttt{nibackgen3C50} tool.
Fig. 6 shows the light curve observed in the 0.2-12 keV range of NICER.
We could not conclude anything definite about the orbital modulation,
 although the variability observed over 35 hours, namely
 the apparent modulation in Fig.6,  supports this possibility.
 Fig. 7 also shows that, even if there are spectral variations
 within hours, there have not been major differences on time
 scales of hours ({\sl NICER-NICER} comparison, panel on the left)
 or of years (panel on the right,
 with the {\sl Chandra} HETG and {\sl XMM-Newton}
 RGS grating spectra compared to the spectrum observed
 with {\sl NICER}). There were some small,
 but exceeding the cross-calibration uncertainties, flux differences 
 between {\sl NICER} exposures, and between the 
 average spectrum observed in 2000 and in 2021.
 These variations were both in  soft and the hard tail of the spectrum,
 so they do not 
 seem to be due to variable, intrinsic absorption. 
\section{V1674 Her (Nova Her 2021): a magnetic nova}
This nova was discovered as the transient TCP J18573095+1653396
 on 2021 June 12.537 UT at mag. 8.4 by Seiji Ueda (Kushiro, Hokkaido, Japan).
 \citet{Munari2021} reported optical spectroscopic confirmation
 that this was indeed a nova in outburst.
The optical decay from optical maximum - around V=6.2 - is quantified as t$_2$ (time for a decay by 2 mag) and was as short as $\simeq$2.5 days.
Few novae have such a rapid decrease from peak optical luminosity, 
indicating a very energetic phenomenon.
  Like YZ Ret, V1674  Her in outburst
was detected at all wavelengths from IR to gamma rays 
\citep{Li2021, Aydi2020}.
 The nova was observed in 2021 June with Swift and the XRT telescope
 did not detect a source between 2021 June 13 and 2021 June 18. After 
almost a week after the optical maximum, on 2021 June 18 a hard X-ray source
 appeared, followed on 2021 July 1st by the initial rise of a luminous supersoft
X-ray source.

 A modulation with a periodicity of 501.4277$\pm$0.0002 s was discovered in 
archival Zwicky Transient Facility images by \citet{Mroz2021, Drake2021}
 and it was followed by the discovery that a very
 close period is measurable in the supersoft X-ray source that 
 appeared shortly after the outburst. \citet{Maccarone2021} reported a modulation
 with a  503.9 s period in a Chandra observation done for 10 ks on 2021 July 10 with
 the HRC-S camera, and in \citet{Pei2021b} we
 reported a modulation with a 501.8$\pm$0.7 s period in the initial NICER data.   
A second Chandra observation by \citet{Drake2021} revealed a period
 of 501.72$\pm$0.11 s in the 0-order light curve of
 an exposure of 30 ks done with the Chandra HRC-S camera and Low Energy Transmission Grating.
 The period measured at optical wavelengths 
 before the outburst is not consistent, within the error estimate,
 with the one measured with Chandra by \citet{Drake2021},
 or even with a new, recent optical period of 501.516$\pm$0.018 s measured by 
\citet{Patterson2021}. Assuming the period is the one
 of the WD rotation, this is an indication of spin down due to 
 ejected mass in the range 0.2-2 $\times 10^{-4}$ M$_\odot$.
 In a paper in preparation (Dobrotka et al.
2022) a group of us will re-examine the Chandra lightcurves and
 try to determine whether, given the superimposed aperiodic variability
 and the non-sinusoidal shape of the modulation, the conclusion
 of a period change and/or different optical and X-ray periods can be
 confirmed. However, here we examine only the NICER data. 

The fact that the period detected at optical wavelength at quiescence is very close
 to the one detected in supersoft X-rays in outburst has been interpreted as
 an indication that Nova Her may host a magnetized white dwarf,
 in a intermediate polar (IP) system.  These systems host WDs whose magnetic field strengths reach
several 10$^5$ Gauss. An accretion disk forms, but it is truncated
where the ram pressure of the matter in the disk is
equal to the magnetic pressure of the WD's magnetic field
(at the Alfv\`en radius) and the matter is channeled to
the poles. WDs in IPs are not synchronized with the orbital
period like the more magnetized polars, and since the magnetic
axes of the WD is generally inclined with respect to their
rotation axes, the asynchronous primary is an oblique rotator.
X-ray flux modulation with the WD spin period is one
of the main observational properties,
 considered the best indirect proof of the IP nature in many systems in which
direct measurements of circular polarization with
 current instrumentation yield only upper limits
for the magnetic field.  The cause of the short (around a minute)
 period modulation of the supersoft X-rays in outburst 
 is not fully understood, but the modulation has been attributed to the WD
 rotation in three other recent novae in outburst, with periods of several
 minutes, detected again at quiescence, along with typical signatures of IPs:
 V4743 Sgr \citep{Ness2003, Leibowitz2006, Dobrotka2017}, V2491 Cyg \citep{Ness2011,
 Zemko2015, Zemko2018} and V407 Lup \citep[][and Orio et al. 2021, in preparation]{Aydi2018}.
 As described above for one of the proposed explanations for CAL 83,
 the modulation of the supersoft X-rays indicates non-homogeneous temperature on the surface of the WD, 
 which presumably is hotter at the poles. 

%
\subsection{The NICER observations of V1674 Her and their timing analysis}
  V1674 Her was monitored with {\sl NICER} on 45 different dates between
 2021/07/10 and 2021/08/31, with daily frequency whenever it was possible.
 The X-ray light curve was previously followed with the {\sl Swift}
 XRT, and although there always was a detectable X-ray source
 from the third day after optical maximum,
 initially there was a faint, non-supersoft X-ray source, most likely
 due to shocks in the ejected matter \citep[see][]{Drake2021}.
 
 We started the {\sl NICER} monitoring only after the luminous supersoft X-ray source 
 was observed to rise with {\sl Swift}. The data were
 again reduced as described for CAL 83, and the light curve we measured is shown in Fig. 8.
 The {\sl NICER} exposures were not always
 continuous,  and the total exposure time in one day varied 
from 476 s (not sufficiently long for period detection) to 15782 s. 
 The modulation with the $\simeq$501 s period is very clear throughout
 the whole period of NICER observations, including the late
 exposures when the count rate had decreased significantly (interval
 J in Fig. 8).
 However, there was also irregular variability almost every day, on time scales
 of a few seconds, and Figs. 9 and 10 show two examples of
 variability observed only within 24 hours in each case.

 In Fig. 11, we show periodograms of the NICER exposures done
 in the time intervals marked from A to J in Fig. 8, namely from 
 2021 July 11 until 2021 July 12, from 2021 July 13 until 2021 July 22,
 from July22 to August 1, from August 1 to August 8, and from
 that day until August 31st.  The peak frequency, listed in Table 6,
  is always clearly detected, until the end of
 the NICER observations, and always falls within the error derived
 in each interval. Because of the spacing between the single short exposures, in some
 cases covering even only one whole period, the plots show several aliases, but
 it is clear that the main frequency peak, corresponding to an average period 
 501.535 s, is well detected and unique. The peak
 frequency is  always the same within a small uncertainty that
 we estimated by fitting a Gaussian to the peak of the power spectrum, 
 and is also listed in Table 6. The derived uncertainty in the period
 was only -0.300 s when the signal-to-noise is the highest, and 
 became -1.330/+1.340 s when the source was fading.   
 Although the Gaussian fitting method gives a rather conservative
 estimate of the error compared with statistical methods,
 given the superimposed aperiodic variability we preferred it,
 and in this case we did not perform a Fisher randomization test.
 We concluded that the NICER data do not support the strong conclusions on
 the spin-down derived in \citet{Drake2021}. 
\begin{figure}
\begin{center}
\includegraphics[width=140mm]{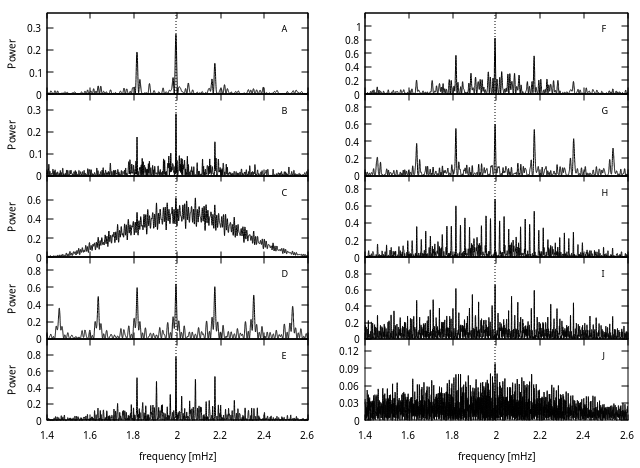}
\end{center}
\caption{The top panel shows the periodograms of the summed exposures 
of V1674 Her
done since the beginning of the observations in the periods marked in Fig. 8. 
 }
\end{figure}
\begin{table}
\caption{Frequency peak in the different time intervals shown in Fig. 8
 in the light curve of V1674 Her.}
\begin{tabular}{cc}
\hline
 Time interval & Frequency peak (mHz) \\
\hline
 A      & 1.9938$\pm$0.0031 \\
 B      & 1.9939$\pm$0.0012 \\
 C      & 1.9935$\pm$0.0044 \\
 D      & 1.9939$\pm$0.0033 \\
 E      & 1.9939$\pm$0.0012 \\
 F      & 1.9937$\pm$0.0021 \\
 G      & 1.9940$\pm$0.0024 \\
 H      & 1.9939$\pm$0.0009 \\
 I      & 1.9939$\pm$0.0006 \\
 J      & 1.9939$\pm$0.0004 \\
\hline
\end{tabular}
\end{table}
 As the light curve in Fig. 8 shows, the total unabsorbed
 average flux decreased since the beginning of August, and 
 the source was observed with NICER until August 31, when the count rate
 had decreased to 1.385$\pm$0.059 cts s$^{-1}$.  
 New observations with the {\sl Swift} X-ray telescope (XRT) were
 done on September 9, 26-27, and October 10-15, still detecting the SSS
 with about constant flux between August 30 and October 16, 
 and with a very similar
 X-ray spectral shape (see Fig. 15).  The period was detected until the end of
 the NICER monitoring. We do not know whether the present X-ray flux
 is representative of the quiescent level over the next few
 years, because there may still be accelerated accretion due to irradiation,
 but since the quiescent IPs X-ray luminosity is
 on average larger than 10$^{32}$ erg s$^{-1}$, at a distance of $\simeq$5 kpc
 \citep[see discussion][and references therein]{Drake2021}
 we expect that V1674 Her will still be detectable 
in a few months after the nova has returned
 to quiescence, and
 detecting the period in the accreting quiescent source will be an
 important test of its IP nature.
\subsection{The X-ray spectral evolution of Nova her 2021}
  \citet{Drake2021} do not discuss spectral
 models that fits the Chandra LETG spectrum
 observed on 2021 July 19  by their collaboration, limiting the analysis
 to identifying spectral lines. However, they also present {\sl Swift}
 XRT monitoring and examine simple models 
 to fit the {\sl Swift} X-ray spectra of the nova. They find a fit
 with a blackbody with four added absorption edges
 (assumed to be of N VI at 0.55 keV,
 N VII at 0.67 keV, O VII at 0.74 keV and O VIII at 0.87 keV
  to simulate typical edges observed in
 such a hot atmosphere), or with the model grid we used above
 for CAL 83 and solar abundances, specifically
 the rauch-H-Ca-solar-90.fits model. 
 They suggest a constant column density N(H)=2.9 $\times$ 10$^{-21}$ cm$^{-2}$
 (approximately the value estimated in the direction of the nova,
 without intrinsic contribution)
 and a constant blackbody temperature of about 61 eV between days 18.9 and
 27.7 after the optical maximum. After this period the blackbody
 temperature increased and reached an astounding value
 of 125 eV (145,000 K) on day 44.2, later oscillating between 65 and 85 eV.
 Although we confirm the goodness of these fits to many of the {\sl Swift} data,
 the better S/N, larger calibrated energy range and same
 or better spectral resolution 
 of {\sl NICER} \citep{Prigozhin2016} allow to observe a more 
 structured spectrum than
 the one measured with {\sl Swift}, and  reveal 
 that the models 
 used for {\sl Swift} are not adequate for better quality spectra. 

 For the  average {\sl NICER}
spectra of different exposures, both on the same day and on
 different dates,  we measured a very similar spectral shape.  
 However, in Fig. 12 we show the spectra extracted
 during the first exposure
 done on 2021 July 20 in short ($\simeq$100-150 s) 
intervals around the maxima and minima,
 normalized to the level of count rate in the 0.2-0.4
 cts s$^{-1}$ range of the first of the two maxima.
 We notice some hardening during minima, that appears evident
 when we compare the ``softness ratio'', that for this source
 we defined as the ratio of the
 count rate in the 0.2-0.5 keV divided by the count rate in
 the 0.5-1.0 keV: 0.25 and 0.27 for the two
 maxima, and 2.77 and 2.41 during the minima.
 This difference indicates that the flux decrease
 may be due to column density variations. We repeated the exercise 
 for maxima and minima during other GTIs and exposures, finding that,
 around the time of the minima, we consistently seem to measure a 
 larger ratio of count rate in the 0.5-10 keV range
 to that in the 0.2-0.5 keV range than around the maxima.
 The ``high state'' average spectrum versus ``low state'' 
 average spectrum comparison in Fig. 3 of
 \citet{Drake2021} for the {\sl Chandra} spectrum 
 around minima showed more flux compared to the continuum
 level for emission lines of iron in the 0.776-0.886 keV 
 region, of O VIII at 0.654 keV and O VII at 0.574 keV, and 
of N VII at 0.500 keV,
 than the average spectrum extracted close to the maxima.
 These emission lines are not resolved with {\sl NICER}, so even
 comparing maximum and minimum spectra obtained during the same 
 period cycle does not allow to clarify whether the less soft
 spectrum is due to emission lines that are more prominent
 above the continuum.
 We cannot rule out that the variation is due only to the
  plasma that emits the emission lines, rather than to the
 WD atmospheric flux, which at this temperature
 should have only absorption features \citep{Rauch2010}. 
\begin{figure}
\begin{center}
\includegraphics[width=70mm]{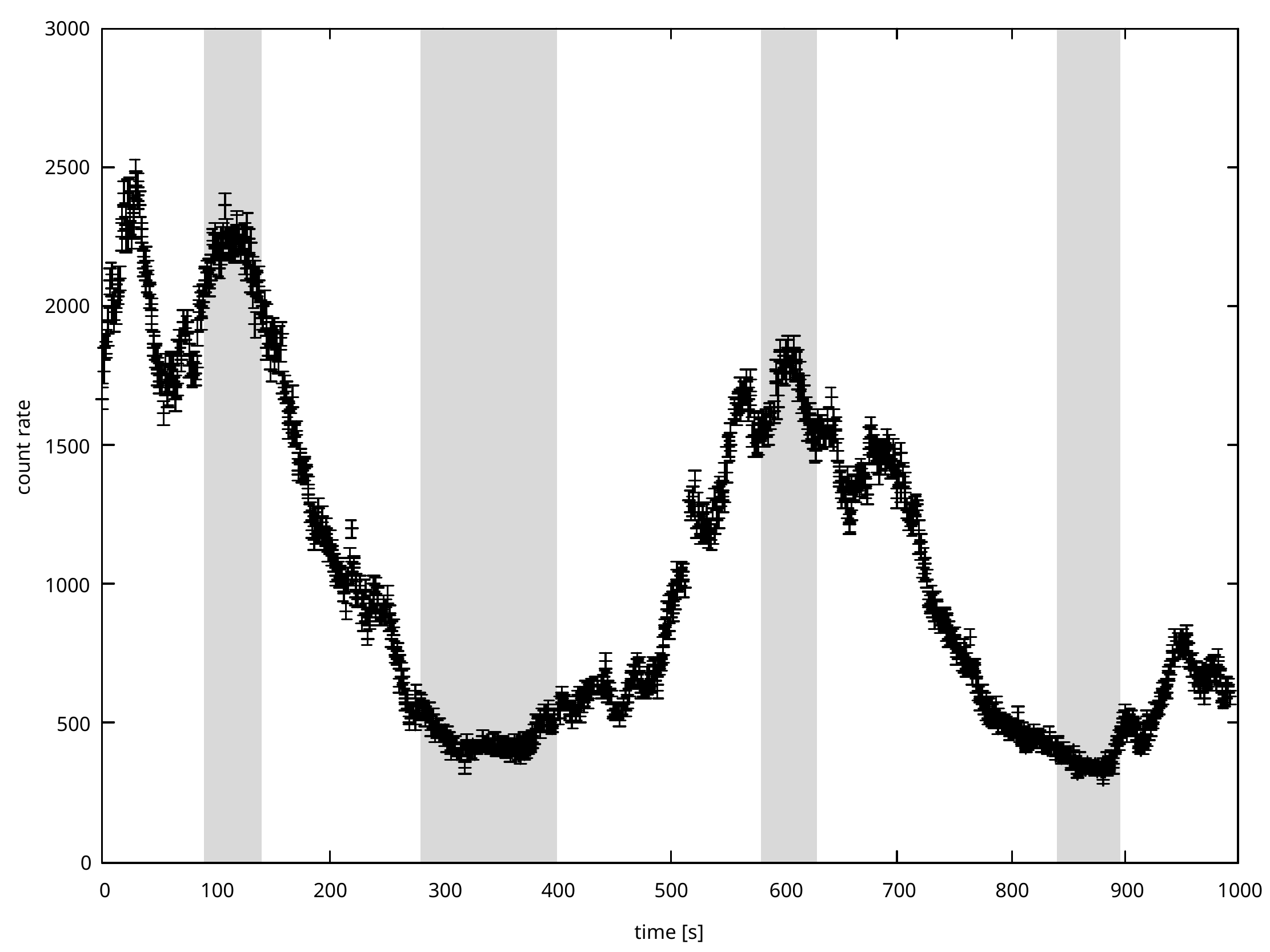}
\includegraphics[width=75mm]{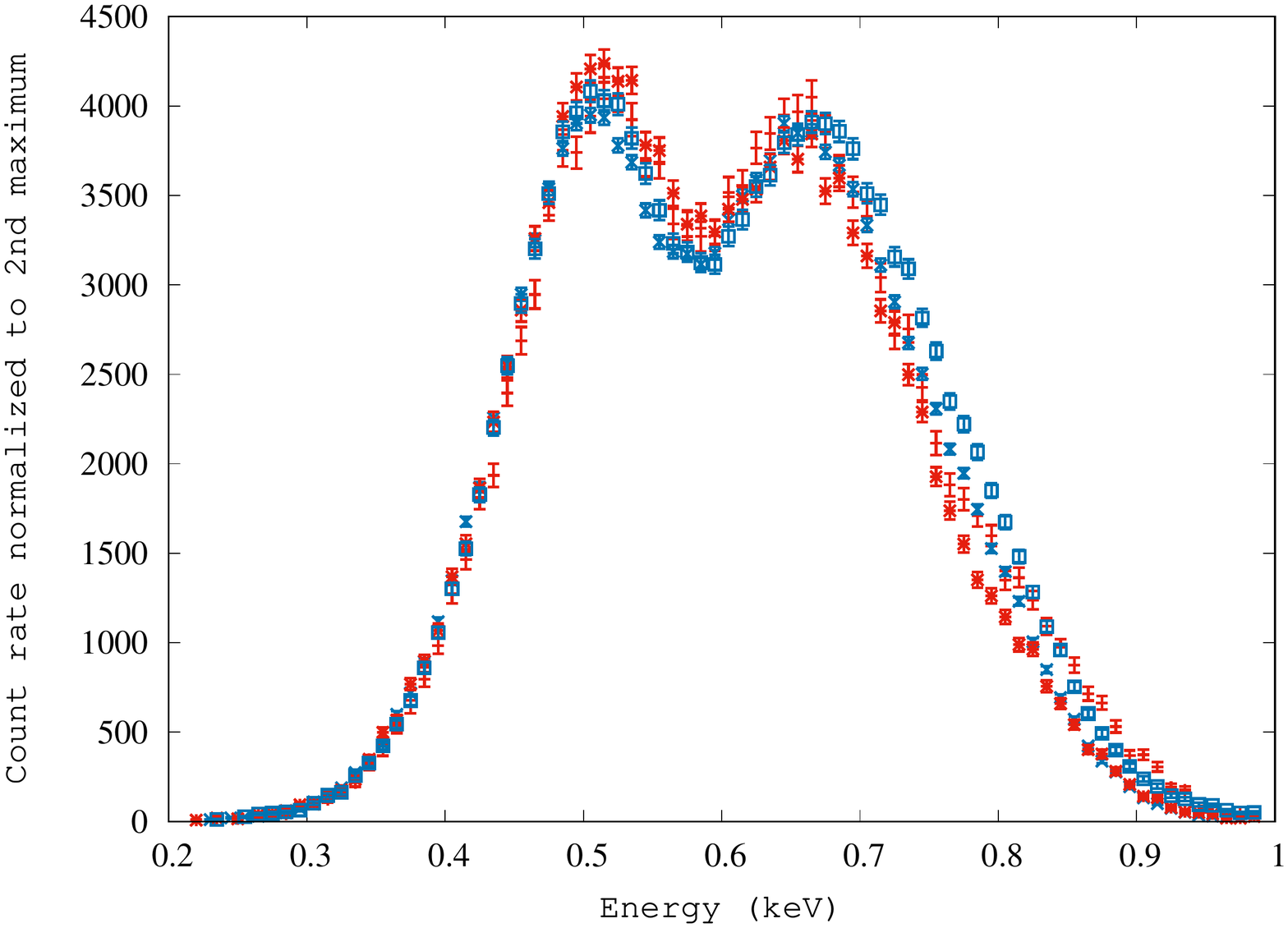}
\end{center}
\caption{The spectra extracted around maxima and minima 
(shaded areas) of the first exposure
 done on July 20 are shown on the right. The spectra traced
 in blue are extracted from the minima shaded on the light curve
 shown on the left, the red  from the minima. The count rate has been 
 normalized to the level of the second maximum, by multiplying the 
 first maximum spectrum (red stars) by 0.7, and the minima respectively
 by 3.2 (first minimum, x-es), and by 2.7 (second minimum, squares).} 
\end{figure}

  We also suggest that the difficulty
 in fitting the data, at least since July 13,  is due to
 these strong emission lines, measurable 
only in the single {\sl Chandra} grating exposure.
 In Fig. 13 we show the {\sl NICER} spectrum observed on 2021 July 19
 and 
the Chandra LETG spectrum observed on the same day,
 which is complex and rich in both emission and absorption
 features.  We also overplot the fit to both  spectra 
 with a blackbody and absorption edges used for the {\sl Swift} spectra 
 of the same period \citep{Drake2021}.
 This is the best fit we obtained to the {\sl NICER} data
  with only one component, but it is not rigorous,
 yielding a reduced $\chi^2$ value larger than 1. 
 The deviation from an atmospheric spectrum due the emission lines
 appears clearly in  the panel on the right, showing the {\sl Chandra}
 spectrum. The prominent, superimposed emission lines, 
 must originate in one or more additional components. 
 We concluded that the blackbody with
 absorption edges is not a significant model for  the SSS phase
 of V1674 Her,
 and that the effective temperature of the WD cannot be estimated in this way.

 Until we have a rigorous model for the {\sl Chandra} LETG spectrum 
 of the supersoft source (which should be the
 subject of an upcoming paper), we cannot draw any conclusions on the
 effective temperature, which is related to the WD mass. 
We note that the atmospheric fits are closer to converge only using
 the grid without enhanced abundances,
 and  at a temperature not exceeding 900,000 K, which
 is significantly lower than the temperature obtained
 with the fits to the {\sl Swift} spectra \citep{Drake2021}
 with blackbody and absorption edges (the column
 density in this case N(H) would be higher).
 However, the only definite conclusion we can draw is 
 not based on the model fits, and is that
 by comparison with other novae's SSS spectra
 \citep[see Fig. 3 of][]{Drake2021} V1674 Her was quite ``harder'' than average,
 indicating a very massive WD. 

 One interesting characteristic of two novae that in recent years have been
 found to be IPs is that the supersoft luminosity emission region
 appeared to shrink rather than cool during the luminosity decay
 \citep{Page2012, Aydi2018}. In 
 Fig. 15 we show that this seems to have been the case also for V1674 Her.
 If the SSS was cooling, it was only by a very small amount, as
 the source flux was decreasing much faster than what may be attributed
 to decreasing temperature.
 A likely interpretation is that the burning ended later at the poles than
 in the rest of the WD, possibly fed by renewed accretion.
 Also in quiescence,  three IP-novae retained a very soft emission region,
 albeit at much lower luminosity, with about the same effective temperature
 as the nova in outburst \citep{Zemko2015, Zemko2016}, Orio et 1l. 2021, in preparation.
\begin{figure}
\begin{center}
\includegraphics[width=87mm]{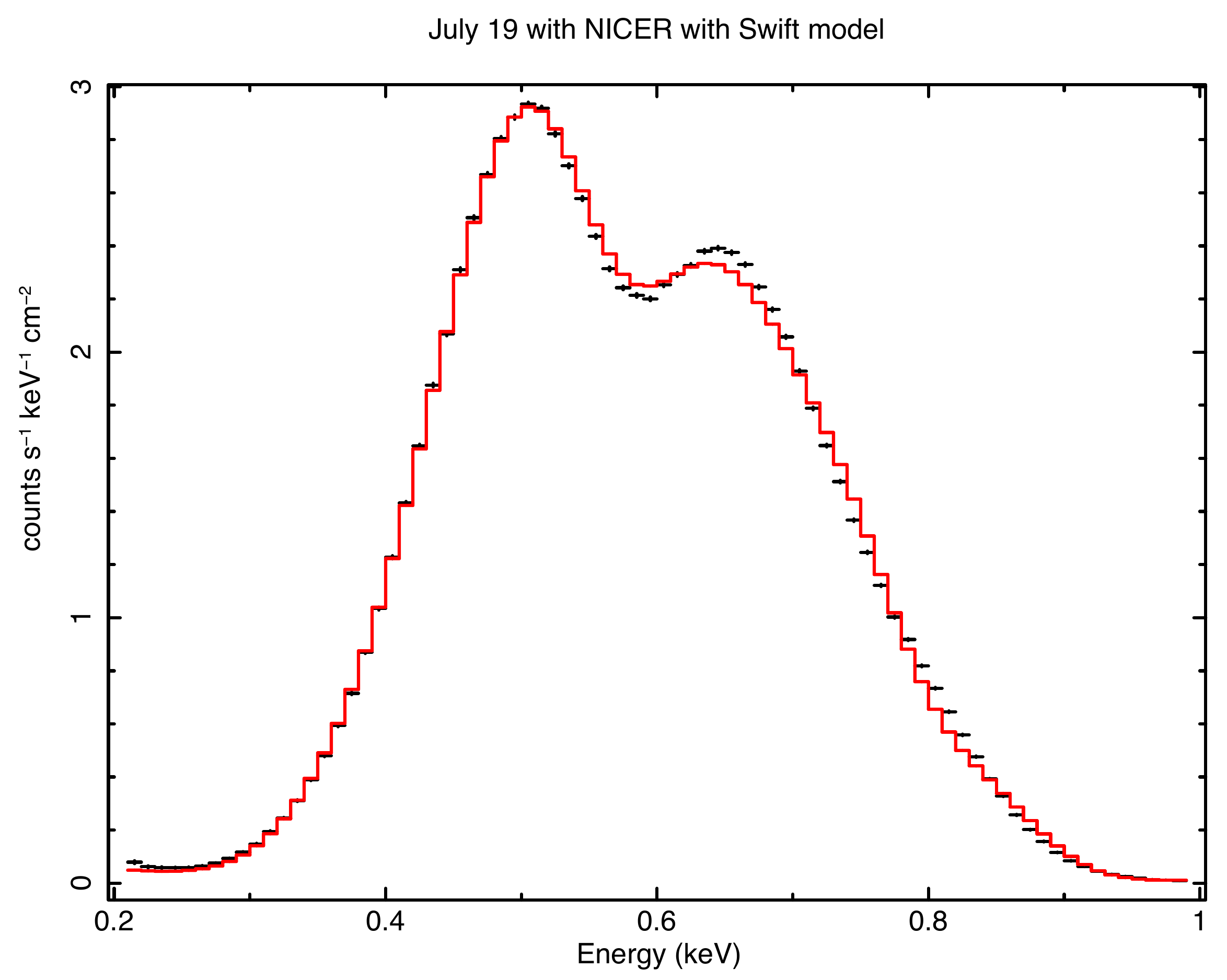}
\includegraphics[width=87mm]{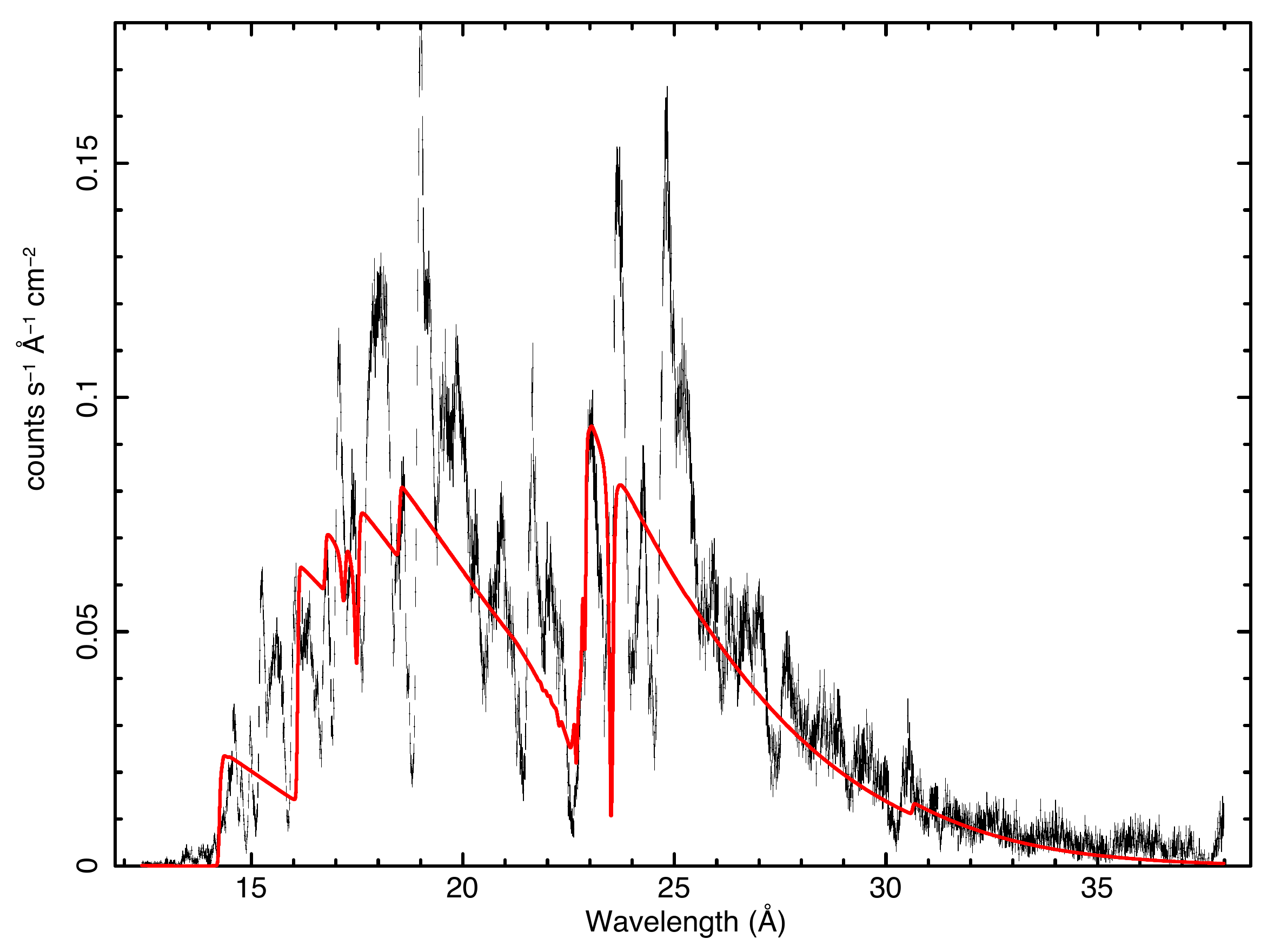}
\end{center}
\caption{The spectrum of V1674 Her on the left, plotted as a function of
 energy, was measured on July 19. 
 The LETG spectrum plotted on the right with the
 natural units of wavelength of the grating was also
 measured on the same day, although at a different time. 
 The best fit with the blackbody+edges model of \citep{Drake2021}
 is shown in both panels. The fit is not perfect for the {\sl
 NICER} spectrum, and since
 in the LETG we are able to resolve the absorption and emission lines, 
 it is clear that  this one-component fit is not adequate. }
\end{figure} 
\begin{figure}
\begin{center}
\includegraphics[width=130mm]{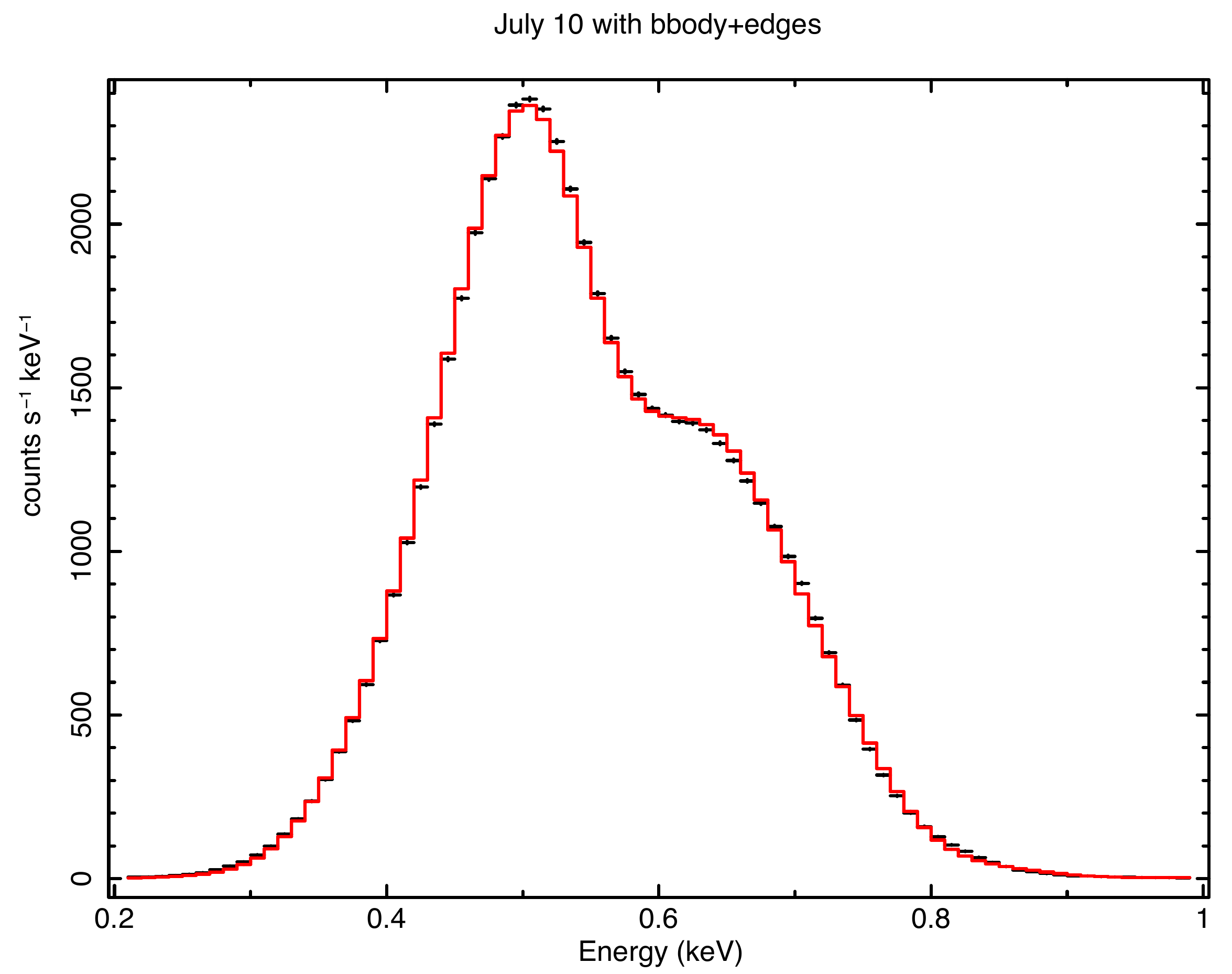}
\end{center}
\caption{The initial {\sl NICER}
 spectrum of of V1674 Her during the bright phase, on July 10, 
 (observations 4622010101), is shown
 here. It is fitted 
 with a blackbody with absorption edges  like in \citet{Drake2021},
 but from July 13 the deviations from a one-component fit
 seem to grow and even increase with time.
 }
\end{figure}
\begin{figure}
\begin{center}
\includegraphics[width=130mm]{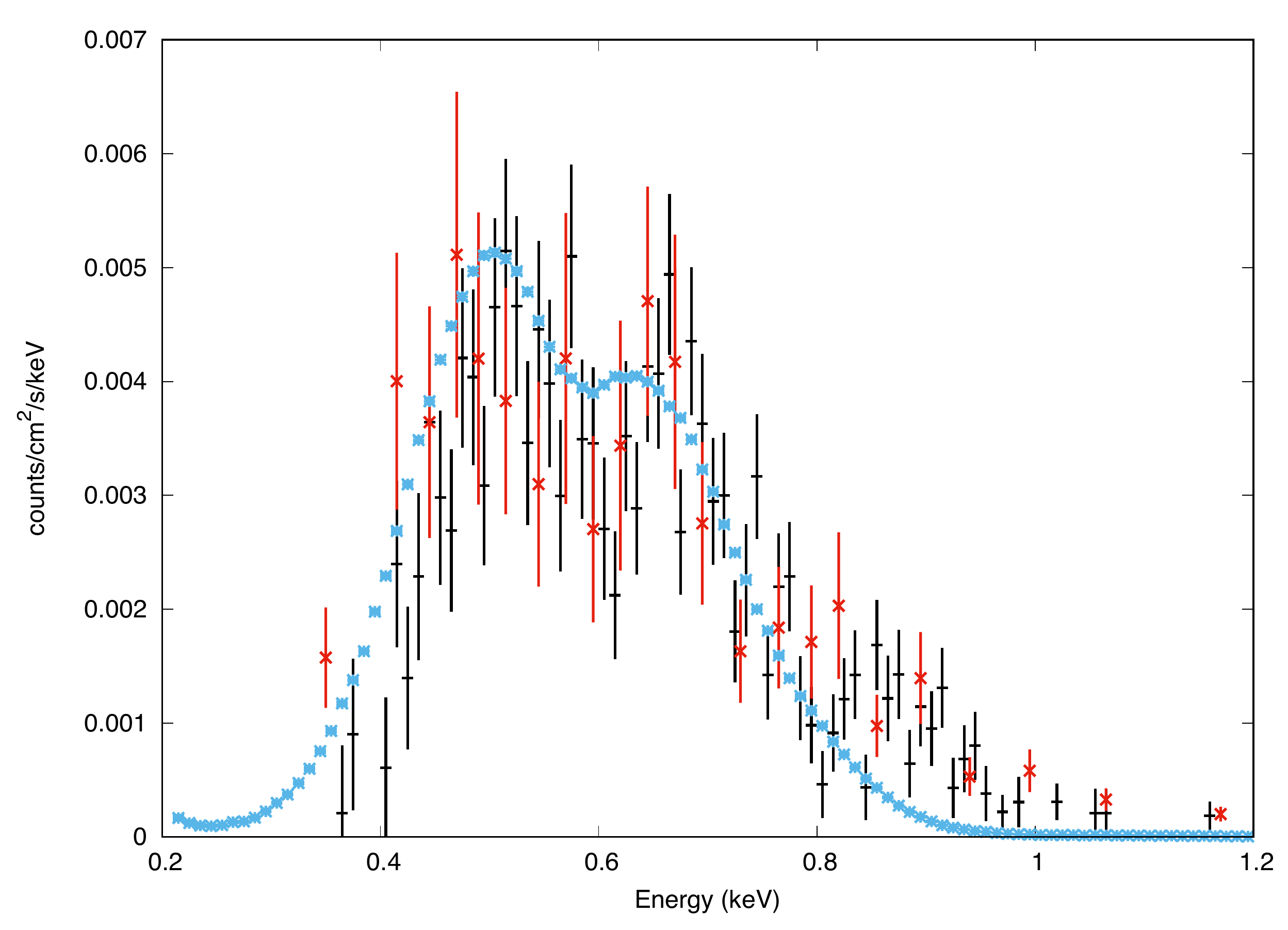}
\end{center}
\caption{Spectra of V1674 Her observed during the decay phase:
the one observed with NICER on 2021 August 3 (observation
 4202260121) is plotted again in light blue divided by a normalization
 factor of 330. The spectrum observed with NICER on 2021 August 30
 (observation 4202260139) 
   is plotted in black and the one observed with the Swift XRT on
 2021 October 15-16 is plotted in red.}
\end{figure} 
\section{YZ Ret: a so far unique ``VY Scl'' nova}
 YZ Ret (Nova Ret 2020, also called EC03572-5455 and MGAB-V207 in different surveys) was discovered in outburst 
 on 2020-07-11.76 at magnitude V$\simeq$5 by Mc Naught (2020, CBET, 4811, 1) 
 but
 the nova was unreported for about a week, because pre-discovery images
 were later examined and the nova was found to have erupted
 on 2020-07-08.171 \citep[see][and references therein]{Sokolovsky2021}.
 It was classified as a nova in eruption thanks to optical spectra
 first obtained by \citet{Aydi2020}.
 Two peculiarities made this nova interesting for our study: in the first
 place, it is the first known VY Scl system to have been observed
 in a nova outburst \citep[see][]{Li2020, Sokolovsky2021}.
 VY Scl are nova-like binaries that undergo transient periods
 of fading of the optical light by 1.5 to 7 mag in less than 150 days,
lasting from weeks to years. In the more common ``high state'', VY Scl
have large optical and UV luminosity, and this is interpreted as
 evidence that most of the time mass
transfer onto the WD occurs at the high rate $\dot m > 10^{-10}$
M$_\odot$ yr$^{-1}$, to sustain an accretion disc in a stable hot state in which
dwarf novae outbursts are suppressed.  The low states have been
attributed to a sudden drop of $\dot m$ from the secondary, or even to a
total cessation of mass transfer \citep{King1998, Hessman2000};
perhaps due
 to spots on the surface of the secondary covering the L1 point 
\citep{Livio1994}
 or non-equilibrium of the irradiated atmosphere of the donor
 \citep[see][]{Wu1995}.

Monitoring the X-rays emission of a few VY Scl WD binaries, 
 no indications have been found of a previously hypothesized
SSS, expected with high $\dot m$ because
 of non-explosive thermonuclear
 burning \citep[see][and references therein]{Zemko2014}.
  At least for one VY Scl, V794 Aql,
 there is no clear correlation between optical and X-ray luminosity
\citep{Sun2020}.
 Mass transfer may be very irregular or sporadic for these systems,
 due to a complicated interplay between the two binary components affecting the
 thermal state of the donor.
YZ Ret has proven that even if thermonuclear burning does occur,
 it must last for quite some time and end in a nova outburst,
 instead of being steady and without outflows as expected for SNe Ia 
progenitors \citep[e.g. models by][]{vandenheuvel1992}.

  The other peculiarity of this nova is that, unlike most Galactic novae which
 are often concentrated towards the Galactic bulge, it is located 
 away from the Galactic plane, with a very low column density 
of only about 1.2-1.3 $\times 10^{20}$ cm$^{-2}$. 
\subsection{NICER observations of YZ Ret}
 We observed the nova with NICER in 20 intervals of about 1000 s each, during an overall time of 29.71 hours starting on 2020 September 28 at 15:25:44 UT.
 The light curve over all this period of observation is shown in
 Fig. 16.  No periodic modulation could be detected in the NICER light curve.
 
 In \citet{Pei2020} we announced the observation and measurement of the
 X-ray source. We found that, since the column density was known
 and the distance determined with the GAIA parallax is
 2.7$^{+0.4}_{-0.3}$ kpc \citep{Bailer2018}, 
 an attempt to fit the spectrum with an atmospheric model 
  yielded a relatively poor fit, but was sufficient to
 obtain an order of magnitude of the X-ray flux.
 We found that the X-ray  
 luminosity was too low to be originating from all, of most of,
 the surface of a hydrogen
 burning WD, only about 2.5 $\times 10^{35}$ erg s$^{-1}$ (see Fig. 17).
 We confirm this result: the closest 
 fit obtained with a model atmosphere (nova model
 SSS$_{-}$003$_{-}$00010-00060.bin$_{-}$0.002$_{-}$9.00.fits
 in the grid by Rauch, see above) yields a reduced $\chi^2$ of 1.37 with
 75 degrees of freedom, with T$_{\rm eff}$=534,000 K and N(H)=3.3 $\times 10^{20}$ cm$^{-2}$.
 The bolometric luminosity would be only
  $\simeq 3 \times 10^{35}$ erg s$^{-1}$ at  distance of 2.7 kpc.

 An observation with {\sl XMM-Newton}, including use of the RGS grating
 and done 5 days earlier,
 explains the low flux: the system must be at high inclination, because 
 no stellar continuum was detected. The spectrum was an emission
 line one \citep{Sokolovsky2021}, probably due
 to photoionized ejecta and with no spectral signatures
 of shocked material in collisional ionization
 equilibrium. Although this conclusion cannot be reached by
 examining our {\sl NICER} broad band spectrum, the bolometric luminosity
 derived is of order 1/1000 of that expected for a burning
 WD atmosphere, about the same as in the
 XMM-Newton observation, so clearly this does not support the possibility that
 the spectrum changed and the central WD SSS spectrum
 emerged during the 5 days intercurring between our {\sl NICER}
 exposures and the data taken by \citet{Sokolovsky2021}.
In a different project collaboration, some of us observed
 the nova after another month with the {\sl Chandra} LETG, and although
 the spectrum had considerably  changed and softened
 in comparison with the observations reported by \citet{Sokolovsky2021}
 and with our {\sl NICER} broad band spectra,
 it was, once again, an emission line
 one, without a measurable stellar continuum \citep[][and Orio et al.
 2022, in preparation]{Drake2020}.

\begin{figure}
\begin{center}
\includegraphics[width=130mm]{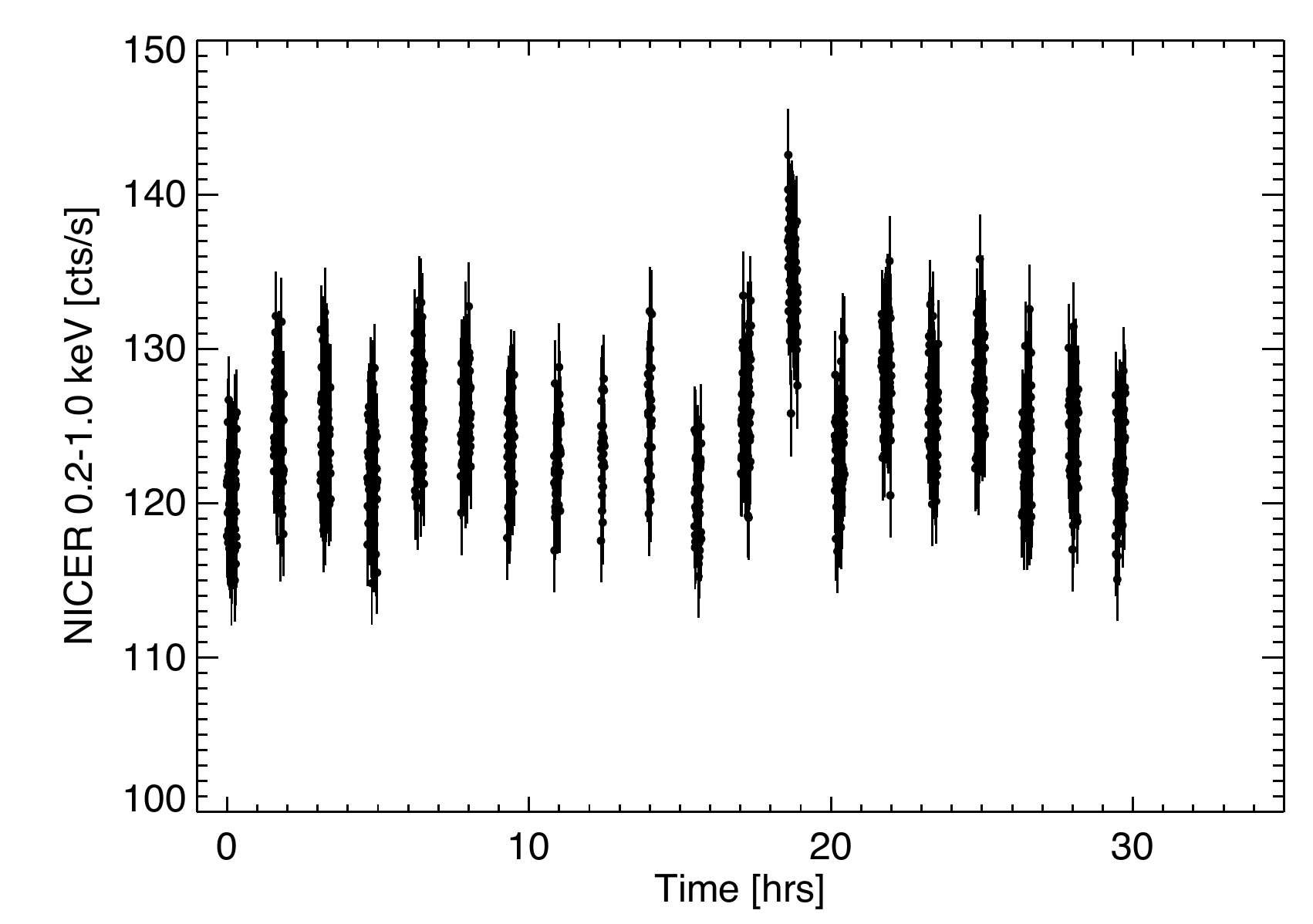}
\end{center}
\caption{Light curve of YZ Ret observed with NICER
 with 16 s time bins, on 2021 September 28 and 29.}
\end{figure}
\begin{figure}
\begin{center}
\includegraphics[width=83mm]{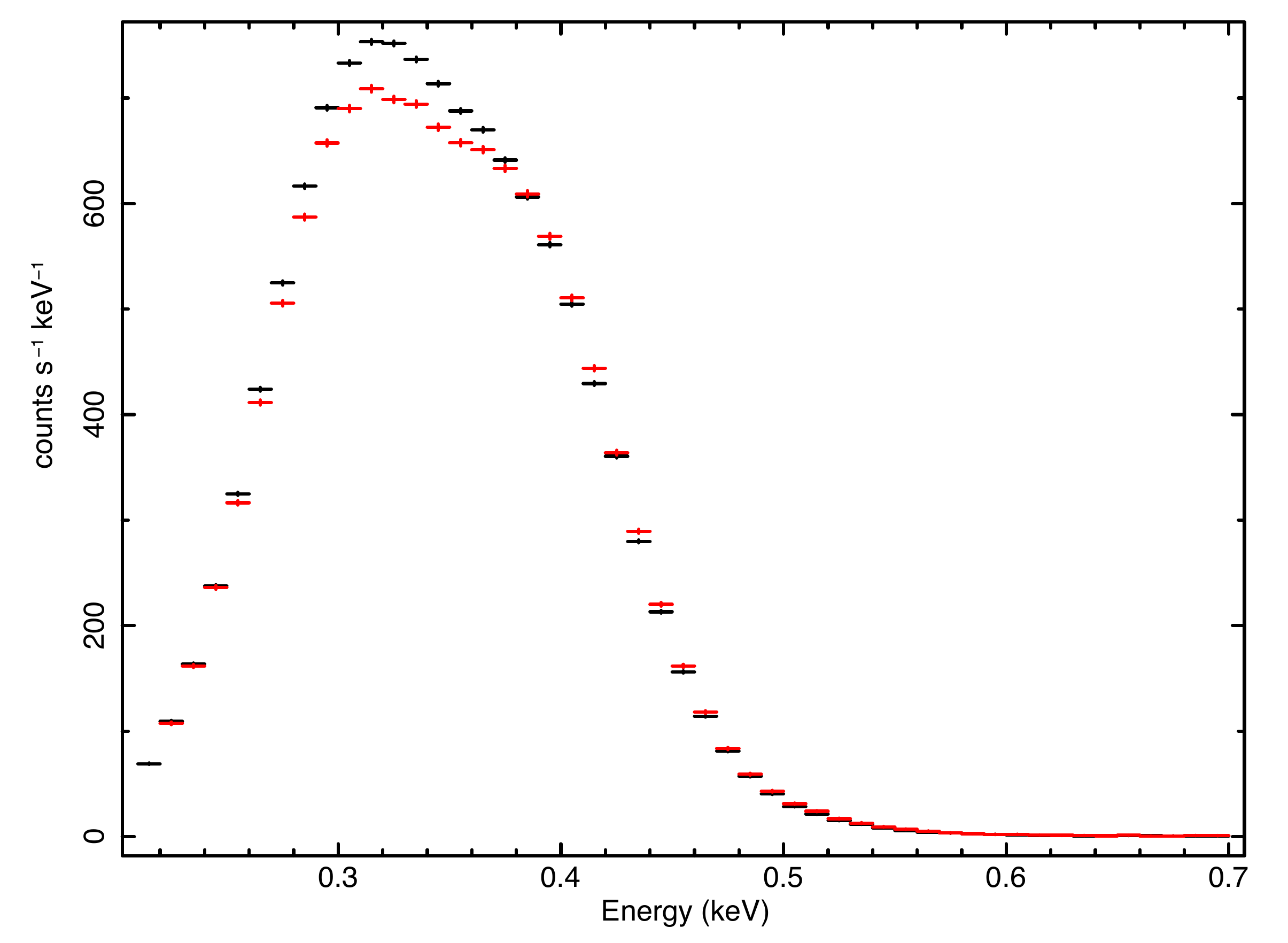}
\includegraphics[width=83mm]{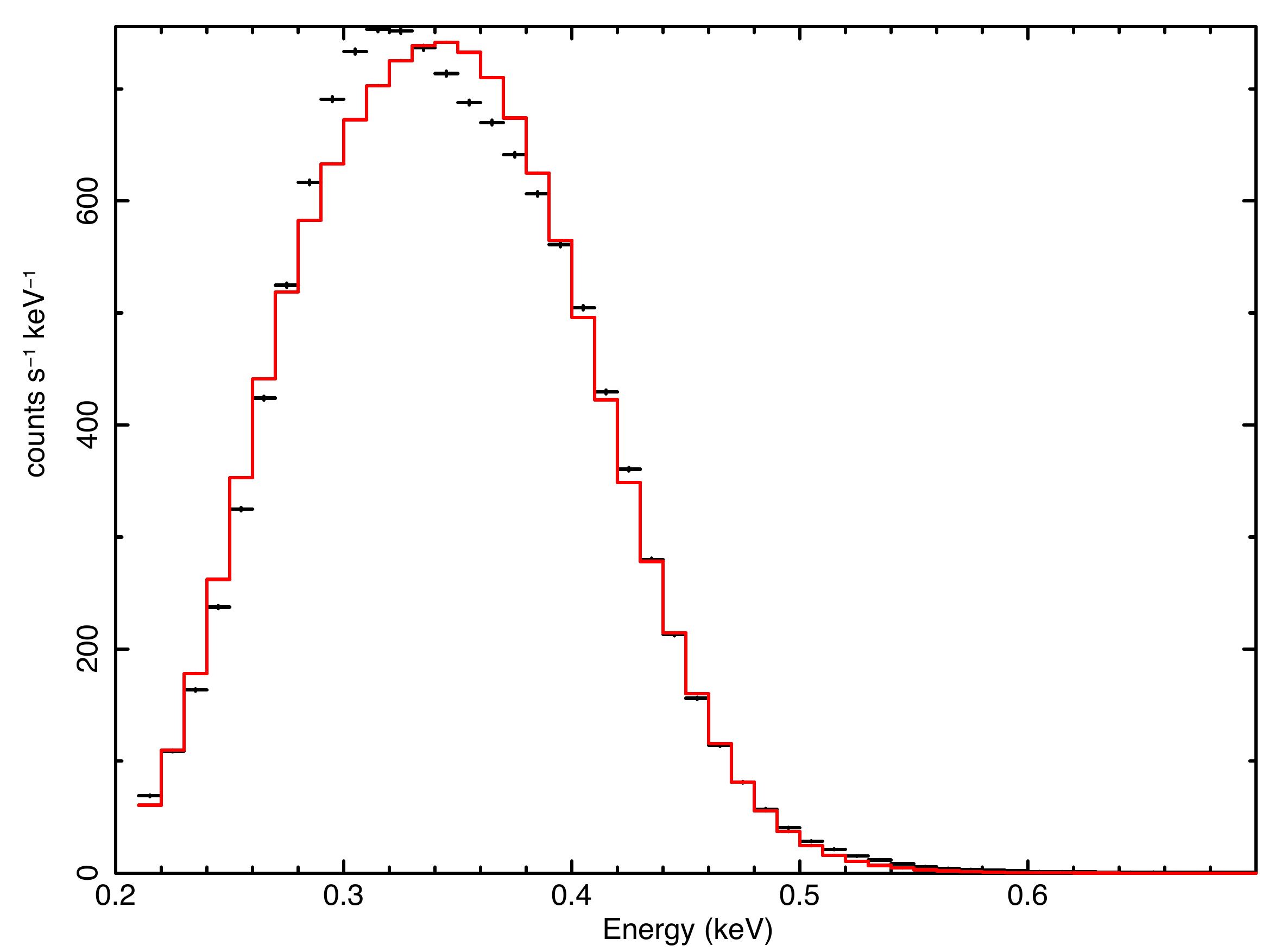}
\end{center}
\caption{Spectra of YZ Ret observed on 2020 September 28 (in black)
 and on 2020 September 29 (in red) on the left, and on the right
 the 29 September spectrum fitted with an atmospheric model (see text).
 We did not obtain a rigorous fit, but we could at least conclude 
that the X-ray flux is orders 
 of magnitude lower than that of a shell burning WD.}
\end{figure}
\section{Conclusions}
We have observed four SSS with NICER, two persistent sources in close binaries
 and two transient ones (two novae). The types of objects
 we observed span almost all the types of WD binaries. 
 The two close binaries appear to be, respectively, a CV-like binary with possibly 
 an unusually massive donor, clearly observed at low inclination (CAL 83), and a
 short-period symbiotic with a yellow donor (MR Vel). 
 The two novae are very likely to be also an IP (V167Her)
 and a VY Scl nova-like (YZ Ret).

$\bullet$ The
 two SSS non-nova close  binaries,
 specifically MR Vel, and CAL 83 in its most common ``high'' state,
 had almost unvaried SSS spectra  since the last
 observation performed 21 and 13 years ago, respectively.
 MR Vel was observed with almost the same X-ray flux. Although
 the flux of CAL 83 varies and there are occasional minima
 when the SSS becomes undetectable, the maximum flux
 is {\bf also about the same}. It is remarkable that
 this source has not been observed to change significantly
 over 40 years of X-ray observations.

$\bullet$ The large deviations from the continuum of a WD atmosphere and
 especially the orders of magnitude lower SSS flux of YZ Ret indicates that
 the WD was never observed in this nova. This is understood if the
 nova system is observed at high inclination \citep[see][]{Ness2013}.
 In agreement with the detection of an emission line spectrum
 with X-ray gratings close to the time of the NICER observations and even
 much later, we attribute the SSS flux to the ejecta,
 which may have either been shocked or photoionized. 
 Given the repeated observations with different 
 satellites at different post-outburst times, we rule out with confidence
 that a very short-lived SSS phase with WD emission was ever
observable for this nova.

$\bullet$ The $\simeq$67 s period of CAL 83 is
 only measurable in X-rays and it has an irregular drift in length,
 by  about 3\%. Our simulations and a rigorous statistical analysis
 were not conclusive using the short {\sl NICER} exposures,
but by going back to the most recent {\sl XMM-Newton} exposure,
 which was much longer and without interruptions,
we found that this drift appears to be real
 and not an artifact of varying amplitude, although the amplitude
 of the modulations is also often observed to vary.

$\bullet$  We confirm the evidence pointing to rotation of a magnetic WD
 as the cause of the modulation of V1674 Her, although in outburst the X-ray
 luminosity was not due to accretion onto the pole, and
 only originated in shell burning. Thus, the WD superficial temperature
 may not have been homogeneous during the SSS phase,
 explaining also the difficulty in obtaining a unique and rigorous
 spectral fit.

$\bullet$ The flux of V1674 Her had irregular fluctuations superimposed
 on the regular modulation, confirming the aperiodic variability detected 
 in other nova SSS (e.g. N SMC 2016, \cite{Orio2018}, and N LMC 2009 \cite{Orio2021}).

$\bullet$ 
 Because of superimposed emission lines - revealed by a {\sl Chandra}
 high resolution spectrum - we were not able to obtain a rigorous fit to
 most of the {\sl NICER} spectra.
 Although the SSS  spectrum of V1674 Her was
 ``harder'' than the SSS spectra of other novae, with
 the higher S/N and better supersoft response of {\sl NICER} we  could
not confirm the extremely high peak temperature of $\simeq$1,450,000 K 
suggested by fitting {\sl Swift} spectra
 \citep{Drake2021}. 

$\bullet$ Comparing the V1674 Her count rate and spectrum around maximum and
 minimum of the $\simeq$501 s period, we find that the soft 
flux is not more absorbed at minimum, so the flux periodic
 variation is not due to periodically varying absorption. 
 The interpretation we suggest is that
 there is a ``SSS-obscure'' region, that has much
 lower effective temperature and does not emit X-rays, possibly around the
 equator, with a sharp gradient between the SSS and non-SSS regions. 
 The SSS regions must occupy a large portion of the surface,  because
 the SSS flux does not disappear during the minima. 
 
$\bullet$ The 501 s ($\simeq$8.35 minutes) period in V1674 Her was constant
 (within the precision of our timing analysis) during the post-eruption
 luminous SSS phase, and a very close period was measured in optical
 in quiescence and in outburst. Within the precision of the {\sl NICER}
 data, we could not confirm the putative spin-up of the WD after
 the outburst.

 $\bullet$ The fading of V1674 Her
without significant spectral changes indicates that the
 SSS emitting region shrunk before cooling, a phenomenon suggested
 to occur also in other ``magnetic novae''. 

$\bullet$ We only observed periodic modulations in
two SSS (nova V1674 Her and the non-nova close binary CAL 83)
in which there are clear indications that we observed the WD
 surface, while no such modulations were measured in two
 sources (again, a nova - YZ Ret - and a non-nova - MR Vel)
in which we only observed a wind or an outflow.
  Considering also that the oscillations
 were never observed in other SSS whose
 spectrum is dominated by emission lines,
and whose flux is much lower than in SSS in which the WD atmospheric continuum
 was observed,
 this pinpoints at the root cause of the periodic flux modulations
 as a phenomenon occurring on the WD surface or very close to it.
 Of course, there are also nova and non-nova SSS in which the
 WD was observed, but 
 periodic modulations were never measured \citep[see][who list 18 SSS
 in which they searched for $\simeq$minute long oscillations, finding
 them only in 5 of them, with a duty-cycle even as short as 11\%]{Ness2015}.
 Only  3 out of the 18 sources in this paper (CAL 87, U Sco and V959 Mon) are 
 at high inclination, and only CAL 87 has a X-ray spectrum that is
 completely dominated by emission lines like MR Vel and YZ Ret. However,
 there are a few caveats. First of all, 5 other of those 18 sources show
modulations of the X-ray flux with tens of minutes (see Table 1), bringing the total
 number of X-ray sources with quasi-periodic or periodic SSS modulations
 to 10 out of 18. If the modulations with periods of
 tens of minutes and those with $\simeq$ 1 minute long oscillations
 are mutually exclusive, as in this sample of SSS, this may be a clue to a rotational
 nature of the shorter oscillations, per analogy with
 the longer ones. A parameter to notice is the effective temperature of
 the non-periodic SSS, as estimated in the papers referenced in Table 1.
 We find that on average, they have lower temperature, although
 there is an overlap with the range of temperature of the periodic sources. 
 The lower average temperature of the non-periodic
 SSS means that also their average X-ray luminosity is lower,
 hindering the timing analysis because of the lower S/N.
 Finally, we note that the ``duty cycle'' (the fraction of time
 in which the modulation is measurable) of the two sources observed with
 {\sl NICER} and described in this paper is about 70\% for CAL 83
  and 100\% for V1674 Her, higher than all measurements of \citet{Ness2015}. For
 the recent outburst of RS Oph, in a separate, still unpublished
 paper in preparation by Orio et al. (2022) we found a 100\% duty cycle of a $\simeq$35 s oscillations, compared
to only $\simeq$16\% found in 2006 by \cite{Ness2015}
 with lower S/N data. We can speculate
 that the variable amplitude of the period may make the modulation
 undetectable when the S/N of the measurement is low, the amplitude of the
 fluctuation is small, and the period is short.
 Does the modulation really disappear for part of
 the time and is it really not present in almost half of the SSS
 with an observable WD? We
 would like to argue that the jury is still out, and that {\sl NICER}
 with its high S/N may allow to answer in the next few years, with
 more observations of SSS.

\section*{Acknowledgement}
MO and MG were supported by NASA awards for {\sl NICER} cycles AO1, AO2 and
 AO3. SP was funded by a PhD stipend of the China Scholarship Council.
AD and J. Magdolen were supported by the Slovak grant VEGA 1/0408/20, and by the Operational Programme Research and Innovation for the project: Scientific and Research Centre of Excellence SlovakION for Material and Interdisciplinary Research", ITMS2014+ : 313011W085 co-financed by the European Regional Development Fund. GJML is member of the CIC-CONICET (Argentina) and acknowledges support from grant ANPCYT-PICT 0901/2017. J. Mikolajewska was financed by Polish National Science Centre grants 2015/18/A/ST9/00746
and 2017/27/B/ST9/01940.
%
\bibliographystyle{aasjournal}
\bibliography{biblio}
\label{lastpage}
\end{document}